\newcommand{\MScolor}{} %to get rid of red highlighting

\documentclass[useAMS,usenatbib,usegraphicx]{mn2e} 
\usepackage{natbib, color}
\usepackage{amsmath,amssymb}
\usepackage{aas_macros} 

\title{Statistical properties of dark matter mini-haloes at $z$ $\ge$ 15}
\author[M. Sasaki et al.]{Mei Sasaki,$^{1, 2}$\thanks{E-mail:
sasaki@stud.uni-heidelberg.de (MS)} Paul C. Clark,$^{1}$ Volker Springel,$^{3}$ Ralf S. Klessen$^{1}$ 
\newauthor
and Simon C. O. Glover$^{1}$\\
$^{1}$Universit$\ddot{a}$t Heidelberg, Zentrum f$\ddot{u}$r Astronomie, Institut f$\ddot{u}$r Theoertische Astrophysik, Albert-Ueberle-Strasse 2, 69120 Heidelberg, Germany\\
$^{2}$International Max Planck Research School for Astronomy and Cosmic Physics at the University of Heidelberg (IMPRS-HD)\\
$^{3}$Heidelberger Institut f$\ddot{u}$r Theoretische Studien, Schloss-Wolfsbrunnenweg 35, D-69118 Heidelberg, Germany}

\begin{document}

\date{Submitted 2014 February 10.}

\pagerange{\pageref{firstpage}--\pageref{lastpage}} \pubyear{2013}

\maketitle

\label{firstpage}

\begin{abstract}
  Understanding the formation of the first objects in the universe critically depends on knowing whether the properties of small dark matter structures at high-redshift ($z \ge 15$) are different from their more massive lower-redshift counterparts. To clarify this point, we performed a high-resolution $N$-body simulation of a cosmological volume  $1\,h^{-1}{\rm Mpc}$ comoving on a side, reaching the highest mass resolution to date in this regime. We make precision measurements of various physical properties that characterize dark matter haloes (such as the virial ratio, spin parameter, shape, and formation times, etc.) for the high-redshift ($z \ge 15$) dark matter mini-haloes we find in our simulation, and compare them to literature results and a moderate-resolution comparison run within a cube of side-length $100\,h^{-1}{\rm Mpc}$.  We find that dark matter haloes at high-redshift have a log-normal distribution of the dimensionless spin parameter centered around $\bar{\lambda} \sim 0.03$, similar to their more massive counterparts. They tend to have a small ratio of the length of the shortest axis to the longest axis (sphericity), and are highly prolate. In fact, haloes of given mass that formed recently are the least spherical, have the highest virial ratios, \MScolor{and have the highest spins}.  Interestingly, the formation times of our mini-halos \MScolor{depend} only very weakly on mass, in contrast to more massive objects.  This is expected from the slope of the linear power spectrum of density perturbations at this scale, but despite this difference, dark matter structures at high-redshift share many properties with their much more massive counterparts observed at later times.
  \end{abstract}


\section{Introduction}

Understanding the formation of cosmic structures at all scales has been of central interest in the field of astrophysics for several decades. We \MScolor{now} have a widely accepted cosmological paradigm to describe the universe, known as $\Lambda$CDM, and its basic physical parameters are well determined today \citep{Kom11}. Once this paradigm is fixed, it is conceptually a straight-forward task to follow density perturbations growing under gravity, allowing one to connect small Gaussian density perturbations in the early universe to non-linear dark matter haloes, one of which hosts our Galaxy.

An extensive body of research on studying large-scale structure formation with $N$-body simulations has been accumulated (\citealt{Efs88}, \citealt*{Lac94}, \citealt*{Kat99}, \citealt{Kau99}, \citealt{Spr05}). More recently, very large simulations that targeted the formation of the Milky Way halo, including of the order of a few billion particles in the high-resolution region, were performed and clarified the hierarchical growth process which formed our Galaxy (\citealt{Die08}, \citealt{Spr08}, \citealt{Ish09}, \citealt{Sta09}). However, much less work has been done on small-scale structure formation, where objects with virial radii of order $\sim {\rm kpc}$ are resolved.

Because there is a turnover in the power spectrum of density fluctuations at $k \sim 0.01\, \MScolor{h{\rm \; Mpc^{-1}}}$ comoving, and the slope of the power spectrum asymptotically approaches the critical value of $-3$ at high wave numbers (see Fig.~\ref{PowerS}), the density perturbations that exist in our initial conditions ($1\,h^{-1}{\rm Mpc}$ in size) at $z \sim 100$ are statistically different from those in a box of $100\,h^{-1}{\rm Mpc}$ in size.  Therefore, there is no reason to expect that the collapsed objects forming at these different scales are strictly self-similar. Our goal is to study the small-scale regime of the power spectrum and the properties of the corresponding first dark matter haloes, and to quantify the differences with larger dark matter haloes that form later.

An important additional motivation for studying these small-scale structures lies in Population III (Pop III) star formation. Recent works (\citealt*{Osh07}, \citealt{GaoL07}, \citealt*{Turk09}, \citealt{PCC11}, \citealt{Greif11}, \citealt{Smith11}, and \citealt{Greif12}) show that the properties of Pop III stars (such as their masses or multiplicity) formed in each dark matter halo strongly depend on the physical conditions in the halo, and halo-to-halo differences can be large.  Since cosmological hydrodynamical simulations are computationally expensive, previous calculations have focused on the first or the first few mini-halos to collapse in the cosmological volume of interest. They studied statistical properties of Pop III stars by using a group of realizations of cosmological simulations and typically studied only one halo from each realization. As such, halo-to-halo differences within a single cosmological realization are still largely unknown, and our current understanding of the Pop III star formation process may be biased by our selection of the first collapsing halo. Recent works such as \citet{Hir13}, however, have increased the number of samples significantly. But it is computationally still not possible to model all the haloes in the simulation domain with hydrodynamic simulations.

In our work, we adopt a pure $N$-body simulation to statistically study halo-to-halo variations on scales relevant for Pop III star formation. We model a cubic region of $1\,h^{-1}{\rm Mpc}$ in size using $2048^3$ dark matter particles with a mass of $\sim 9 \,{\rm M}_{\odot}$ each, and follow the dynamical evolution of this region from a redshift of $z \sim 100$ to $z = 15$. Full details of the simulation are provided in Section 2.  \MScolor{To date many numerical studies of Pop III star formation have used a friends-of-friends (FOF) method to identify haloes in the computational volume \citep{Y03}. In order to make a meaningful comparison with these Pop III studies, we adopt a similar approach here. However, it has been shown that such a method of decomposing the dark matter structure can produce noisy results, with particles near the boundaries switching between neighboring haloes somewhat randomly each time step, or `bridging' distinct bound structures into single larger structure \citep{BG91}. We demonstrate in Fig.~\ref{fig1} and Fig.~\ref{fig2} that this is more extreme for the high-redshift dark matter haloes found in our simulations.
To account for this feature of the FOF method, we use the {\small SUBFIND} (\citealt{S01}) method to identify dense regions inside each FOF group, and we use the most massive of these `subhaloes' for analysis. Haloes in which {\small SUBFIND} fails to identify substructure ($< 1$ percent of all haloes) are removed from our analysis.}   

Among the previous works, our study has many similarities with the work of \citet*{Jan01}. We probe slightly higher redshift than they did (their analysis is at $z$ = 10, whereas ours is at $z$ = 15), and our simulation is carried out at significantly higher resolution: we adopt $2048^3$ particles, compared to the $128^3$ particles used in their study, for modelling essentially the same comoving volume in space. As such, our study presents a factor of 4000 improvement in mass resolution compared to previous systematic studies of this kind. There have also been more recent attempts to study the properties of dark matter haloes at $z = 15$ (\citealt{Dav09}, \citealt{Dav10}). While they concentrated on relatively limited properties of dark matter halos (e.g. shape, angular momentum, clustering), our work investigates broader aspects of haloes including formation time and how physical quantities depend on formation time. In short, this paper aims at new precision measurements of statistical properties of dark matter haloes in the early universe as enabled by our high-resolution $N$-body simulations.

This paper is structured as follows. In Section 2, we discuss the numerical methods adopted in our work, and in Section 3, we present and discuss our results. Finally, we give our conclusions in Section 4. An Appendix informs about technical aspects of our analysis such as convergence tests.

\section{Methods}
We have performed an $N$-body cosmological simulation with $2048^3$ dark matter particles, using the {\small GADGET-3} code. Our initial conditions were generated using N-GenIC, the initial condition generator originally developed for the Millenium Simulation \citep{Spr05}. We first identify haloes by the friends-of-friends (FOF) algorithm with a standard linking length of $b = 0.2$ in units of the mean particle spacing, corresponding to $\sim 0.1\,h^{-1}{\rm kpc}$ in comoving units, followed by an application of the substructure finding algorithm {\small SUBFIND} to identify bound structures within haloes. \MScolor{Substructures thus defined are used for constructing a merger tree, and the most massive substructure identified in a given FOF group, \MScolor{referred to as `main subhalo'}, is analyzed throughout this study, unless otherwise noted.}

%\footnote{How can readers get data? - We don't keep as many snapshots as in Millenium simulation, so I think this is impossible. Ralf's idea:state that it is available upon request. Paul suggests that we could ask Sven about setting up a server.}. 

We employ cosmological parameters consistent with the WMAP-7 measurements ($\Omega_m = 0.271$, $\Omega_{\Lambda} = 0.729$, $\sigma_8 = 0.809$, $h = 0.703$, \citealt{Kom11})\footnote{We would not expect our results to differ significantly if we were to use the WMAP-9 parameters \citep{Hin12}, or those measured by Planck \citep{Pla13}.}. The simulation box is $1\,h^{-1}{\rm Mpc}$ in length, and thus the particle mass in our simulation is $\sim 9\,{\rm M}_{\odot}$. We set the gravitational softening length to be $0.01\,h^{-1}{\rm kpc}$. The lengths quoted here are in comoving units. We started from a redshift $z \sim 100$ and followed the formation of dark matter haloes down to $z = 15$, by which time numerous dark matter haloes capable of hosting Pop III stars have formed. Unless otherwise noted, we analyzed the simulation output at the end of simulation (i.e.~$z = 15$). Hereafter we refer to this simulation as the `Small' run. 
  
We also performed a moderate-resolution comparison simulation at larger scales, using $512^3$ particles to follow structure formation inside a region of $100\,h^{-1}{\rm Mpc}$ on a side with force softening length $4.0\,h^{-1}{\rm kpc}$. The results of this comparison simulation are mostly presented in Section \ref{sec_app}, which we will refer to where necessary. This simulation is referred to as the `Large' run hereafter. The essential parameters of the two simulations are summarized in Table~\ref{Mil}.

\subsection{Merger trees}
\label{mtree}

A FOF group is identified by linking all particles that are separated by less than a fraction $b$ = 0.2 of the mean particle separation. Thus, this method is approximately selecting regions that are by a factor of $1/b^3$ denser than the mean cosmic density (by this choice of $b$, naively we expect that the selected region corresponds to an overdensity $\sim 200$ relative to the background, similar to the expected virial overdensity according to the top-hat collapse model). However, FOF is known to occasionally link separate objects across particle bridges, and it is also not suitable for identifying bound substructures inside dense regions. 

Therefore, in each FOF group, bound substructures are identified using {\small SUBFIND}\footnote{We refer the reader to \citet{S01} for the details of the algorithm.}. In short, inside each FOF halo, locally overdense regions are spotted by identifying saddle points in an adaptively smoothed dark matter density field. The latter is constructed with an SPH smoothing kernel with $N_{\rm dens} = 64$ neighbors, while the topological identification of locally overdense regions is based on $N_{\rm ngb} = 20$ nearest neighbors (the notation here follows \citealt{S01}). Two structures connected with only a thin bridge of dark matter particles would be identified as two different (sub)halos by {\small SUBFIND}. Each dark matter particle inside a FOF group is either associated to one subhalo or to none. In the algorithm, all subhaloes are checked to see whether they are gravitationally bound. In defining a formation time for dark matter halo, we shall concentrate on the most massive subhalo in each FOF halo at the final output time; the most massive subhalo typically contains a dominant fraction of the mass of its host FOF halo.

For each FOF group, we investigate when its most massive subhalo gained half of its final mass at $z = 15$. This is done by following the merger tree along the most massive progenitor in adjacent snapshots. We estimate the formation time by linearly interpolating between the bound mass of subhalo at two subsequent {\small SUBFIND} output times: one output immediately after the the subhalo gained half its final mass, and one output immediately before. A similar approach is adopted in works such as that of \citet{Gao05}.

\section{Results}

In the following, we determine and discuss various physical properties that characterize dark matter halos, and try to clarify differences and similarities of dark matter structures that reside at early time in a small box of $\sim 1\,h^{-1}{\rm Mpc}$ on a side, relative to those forming later in a larger ($\sim 100\,h^{-1}{\rm Mpc} $) simulation box.  

%These results are the most precise measurements of such global physical parameters of %dark matter haloes in these small scales to date.

\begin{figure*}
  \begin{center}
    \includegraphics[width=15.0cm]{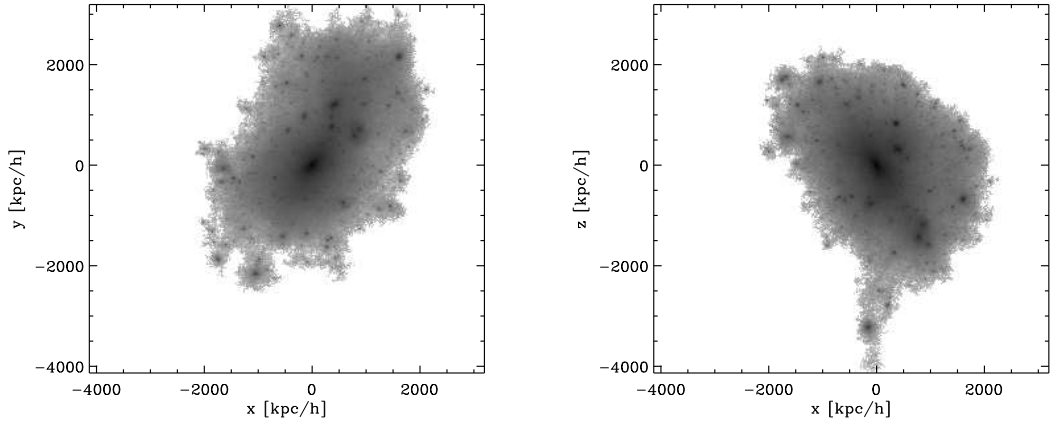}
    \caption{One example of a FOF halo found in the `Large' run at $z = \rm 0$. Its shape resembles a sphere, and it is rich in substructure.}
    \label{fig1}
  \end{center}
\end{figure*}

\begin{figure*}
  \begin{center}
    \includegraphics[width=15.0cm]{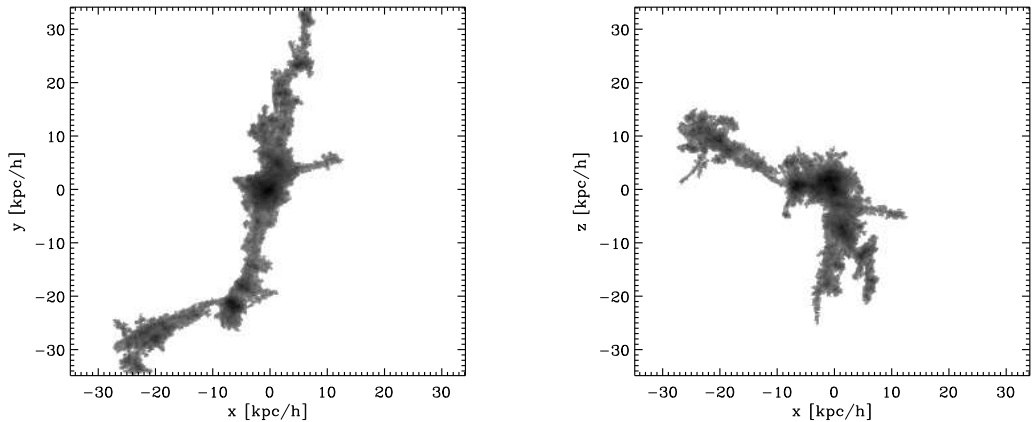}
    \caption{One example of a FOF halo found in the `Small' run at $z = \rm 15$. It has a filamentary shape.}
    \label{fig2}
  \end{center}
\end{figure*}

In Fig.~\ref{fig1} and Fig.~\ref{fig2}, we show representative density maps of FOF objects taken from our `Large' and `Small' runs, respectively. It is easily recognizable that the particular FOF halo shown at $z = 15$ is filamentary and appears to be composed of different dense regions that are connected together. In contrast, the FOF halo at $z = 0$ is much more spherical, and contains rich substructure. While not as extreme in all cases, we confirmed these general differences through inspection of a large number of images of different halos.  
\MScolor{Since the particles located at the outer regions of each FOF haloes are typically not bound to the halo, and could therefore obscure our analysis of the halo binding, shape, etc. in later sections we chose to analyse the most massive substructure identified by {\small SUBFIND} in each FOF halo.}

\subsection{Spin parameter}
\label{sec_spin}
We begin with the dimensionless spin parameter of a halo. It is defined as 
\begin{equation}
  \lambda = \frac{J |E|^{1/2}}{G M^{5/2}},
\end{equation}
where $J$ is the total angular momentum of a halo, $E$ is the total energy, $G$ is the gravitational constant, and $M$ is the mass.  For objects in Keplerian rotation, this value is of the order of unity. For our analysis \MScolor{of angular momentum}, we included only haloes \MScolor{that have at least 1000 particles inside their main subhaloes}, i.e.~with a minimum mass of $\sim 10^4\,{\rm M}_{\odot}$. The spin of dark matter haloes is a result of tidal torques that they experience during their formation and subsequent evolution \citep{Whi84}.

\begin{figure}
\includegraphics[width=8cm]{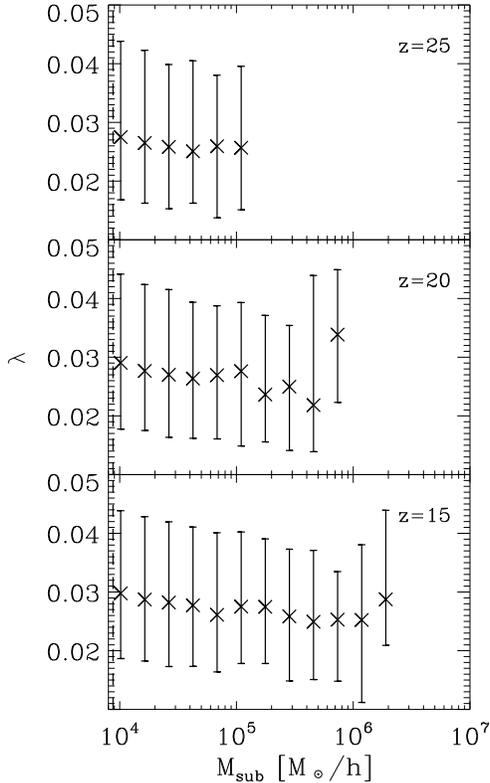}
\caption{Distribution of the dimensionless spin parameter $\lambda$ in dark matter haloes of different mass at redshift $z \sim 25$ (top panel), $z \sim 20$ (middle panel), and $z \sim 15$ (bottom panel). Crosses indicate the median value of $\lambda$, while the error bars indicate the 20th and 80th percentiles. \MScolor{The dashed vertical lines represent lower mass limit of 1000 particles.} We exclude any bins that contained less than 10 samples. The same holds for Fig.~\ref{virial}.}
\label{spin15}
\end{figure}

We show the median, and 20\%, 80\% percentiles of this parameter in a series of logarithmic mass bins in Fig.~\ref{spin15}. \MScolor{We find that at all the redshifts we looked at, the spin parameter $\lambda$ weakly depends on mass.} 
 
\begin{figure} \includegraphics[width=8cm]{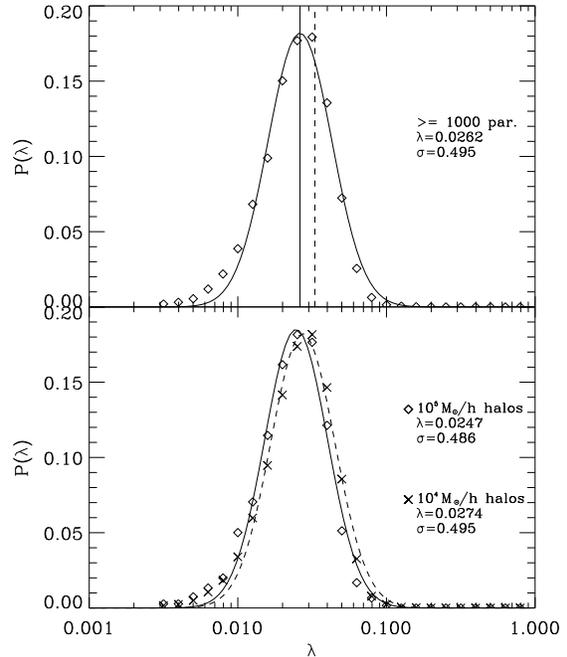} \caption{The top panel is the spin distribution for all haloes with \MScolor{$\geq 1000$} particles at $z = 15$. The bottom panel shows the spin distribution for haloes within a mass range of $10^{5\pm 0.2} \,h^{-1} {\rm M}_{\odot}$ (diamonds) and $10^{4\pm 0.2} \,h^{-1} {\rm M}_{\odot}$ (crosses). The symbols represent data from our simulations, and the lines indicate log-normal fits. The vertical solid (dashed) lines represent $\bar{\lambda}$ for our (\citealt{Jan01}) simulation.}
\label{logall}
\end{figure}

The distribution of $\lambda$ is often fitted by a log-normal distribution \citep*{War92}: \begin{equation}
  p(\lambda)\, {\rm d} \lambda = \frac{1}{\sqrt{2 \pi} \sigma_{\lambda}} \exp \left[ - \frac{\ln^2(\lambda/\bar{\lambda})}{2 \sigma_{\lambda}^2}\right] \frac{{\rm d} \lambda}{ \lambda}. 
\end{equation}
\noindent
We obtained a log-normal fit to the spin distribution at $z = 15$ with parameters $\bar{\lambda} = \MScolor{0.0262}$ and $\sigma_{\lambda} = \MScolor{0.495}$ (Fig.~\ref{logall}, top) for all haloes with at least \MScolor{1000} particles. \MScolor{In other parts of this paper, lower limit of 100 particles is introduced such that lower mass haloes, which are not well-resolved, will be excluded from our analysis in this paper (shape, virial ratios, formation times etc). We adopted a more demanding criterion here, since the spin parameter is known to depend more strongly on how well the dark matter haloes are resolved than other physical parameters (\citealt{Dav10}, see Appendix~\ref{discuss-l} of this paper).}
\citet{Jan01} found that the spin distribution follows a log-normal distribution at $z$ = 10 (with $\bar{\lambda} = 0.033$, which is overplotted with a dashed-line in Fig. \ref{logall}). The similar value in $\bar{\lambda}$ is a bit surprising, because although we model essentially the same volume in comoving space with slightly different redshift, the different mass resolutions in the two simulations mean that we resolve dark matter haloes in different mass ranges. This outcome is however consistent with findings from analytical works that predict that $\bar{\lambda}$ has no strong dependence on the power spectrum \citep{Hea88}. 

\MScolor{Since we calculate spin parameter from substructure within FOF haloes, we do not suffer from artificial high-end tail of spin distribution in Fig.~\ref{logall} as in \citet{Bet07}, who analyzed a simulation output produced with the same code as ours, {\small GADGET-3}, on a larger scale but chose to analyze FOF haloes.}

 Because the spin of a halo is a sum of slight differences in position and velocity space, it is sensitive to how well the halo is resolved. As such, our study qualifies as the most precise measurement of the spin parameter at $z = 15$ thus far. We expect that our estimate of the spin parameter for \MScolor{sub}haloes with mass $> 10^4 \,h^{-1} {\rm M}_{\odot}$ (corresponding to $> 10^3$ particles) is correct within a factor of 2 at a $1\,\sigma$ level \citep{Tren10}. We obtain $\bar{\lambda} = \MScolor{0.0247}$ and $\sigma_{\lambda} = \MScolor{0.486}$ for haloes in the mass range $10^{5\pm 0.2} \,h^{-1} {\rm M}_{\odot}$ and $\bar{\lambda} = \MScolor{0.0274}$ and $\sigma_{\lambda} = \MScolor{0.495}$ for haloes in the mass range $10^{4\pm 0.2} \,h^{-1} {\rm M}_{\odot}$ (Fig.~\ref{logall}, bottom). \MScolor{This is in contrast to the result of \citet{Dav09}, where they find the more massive haloes to have systematically larger spin.} \MScolor{This is a result of different halo identification algorithms. The particles that lie in the surfaces of dark matter haloes contribute little in terms of mass but much in terms of angular momentum. Therefore, the actual value of spin parameter heavily depends on halo identification methods employed. Comparison of Fig.~\ref{spin15} and \ref{fig:FOF} shows that the FOF gives a positive correlation between spin and halo mass, which is not present when we employ the more conservative {\small SUBFIND} method.}

 In simulations with dark matter and gas, it has been found that the spin parameters of both components follow similar distributions \citep{V02}. As we aim to study the environmental conditions of first star formation, knowledge of $\lambda$ is a key prerequisite. There is a recent work that estimates the Pop III initial mass function (IMF) from the rotational velocity of haloes \citep{Souz13}. \citet{Hir13} selected $\sim 100$ haloes from multiple realizations of structure formation and resimulated many samples of Pop III star forming regions in a suite of two-dimensional radiation hydrodynamic simulations, ending up with a statistical study of the final mass of a single protostar. Their spin distribution the of dark matter component is centered on $\bar{\lambda} = 0.0495$, considerably higher than the values found in most pure dark matter simulations including ours. \MScolor{This is largely due to a selection effect and the positive mass-spin relationship of FOF haloes described earlier: these authors employed a standard FOF algorithm to extract the haloes, and focused on the more massive haloes in the simulation (i.e. those massive enough to have triggered $\rm H_2$ cooling), which are likely to have higher spin. }

\subsection{Virial ratios}

In this paper, we define the virial ratio as $2.0 \times {E}_{\rm K} / {E}_{\rm G}$, so that a value of unity corresponds to virial equilibrium. The total kinetic energy ${E}_{\rm K}$ is defined as 
\begin{equation}
{E}_{\rm K}  = \frac{1}{2} \Sigma_i \left. m_i v_i^2 \right., 
\end{equation}
with velocity being defined relative to the center-of-mass velocity of the halo throughout this paper. The gravitational potential energy ${E}_{\rm G}$ is defined as 
\begin{equation}
{E}_{\rm G} = \frac{1}{2}\sum_{i \ne j} G \frac{m_i m_j} {r_{ij}}. 
\end{equation}
The sum is taken over all particles $i$ and $j$ \MScolor{that belongs to the main subhalo in a specific FOF halo}, with $m_i$ and $m_j$ being their mass, and $r_{ij}$ their distance. \MScolor{This quantity is numerically calculated by direct summation.}

\begin{figure}
\includegraphics[width=8cm]{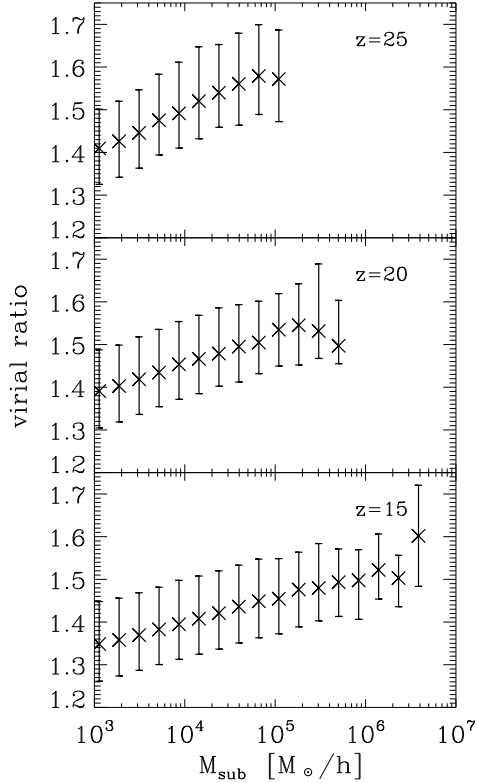} 
\caption{Distribution of the virial ratio in dark matter haloes of different mass at redshift $z \sim 25$ (top panel), $z \sim 20$ (middle panel), and $z \sim 15$ (bottom panel). The virial ratio is defined such that it tends to unity at virial equilibrium. Crosses indicate the median value of the virial ratio, while the error bars indicate the 20th and 80th percentiles. The mass bins are spaced logarithmically, and symbols are only plotted for bins containing at least 10 dark matter haloes.}
\label{virial}
\end{figure}

Fig.~\ref{virial} shows the distribution of virial ratios for different dark matter haloes at redshifts $z \sim 25$ (top), 20 (middle), 15 (bottom). We see that there is some variation in this quantity, but the median, and the 20\%, 80\% percentiles of the virial ratio are well above unity for all mass ranges we have investigated, and thus the dark matter haloes are not virialized but instead are usually perturbed. This has previously been noted by \citet{Jan01} and \citet*{Dav10}, with relatively low resolution at high redshifts, and also by, for example, \citet*{Hetz06} at lower redshifts, $z < 3$. It is also beneficial to compare results of the `Large' box run with the `Small' run. It is clear from Fig.~\ref{Mil-vr} that dark matter halos found in the larger simulation box at $z = 0$ have systematically lower virial ratios, and are thus closer to virial equilibrium than those at $z = 15$.

The median value of the virial ratio in Fig.~\ref{virial} \MScolor{increases} with increasing halo mass. \MScolor{The median value of a given mass bin does not evolve considerably over a range of redshift, in contrast to \citet{Hetz06}, who have found that the virial ratio decreases monotonically with redshift between $z = 3$ and $z = 0$.} These differences reflect the different dynamical states of dark matter mini-haloes at $z \geq 15$ and more massive systems at $z < 3$.

Our results show that dark matter haloes at $z \sim 15$ cannot be considered to typically represent isolated systems undergoing collapse (see also discussion in subsection \ref{sec_trts}). The excess kinetic energy of the dark matter haloes, if shared by their gas component, could influence the star formation taking place because the properties of the turbulence in the interstellar medium strongly influence the star-formation process within them (\citealt{PCC11}, \citealt*{PriJ12}).  Distinguishing relaxed haloes from unrelaxed haloes is not a straight-forward task, and needs to be based on complex criteria that involve, for example, virial ratios and the fraction of mass in substructures \citep{Net07}. Our simple definition of virial ratio should however already give a useful proxy for the dynamical state of a halo.

\subsection{Mass function}

\begin{figure}
  \includegraphics[width=8.1cm]{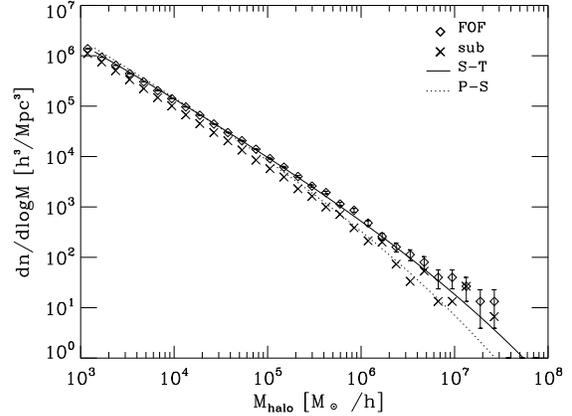}
  \caption{Mass function of dark matter haloes at $z = 15$ \MScolor{(diamond symbols represent number counts for FOF haloes, when crosses denote those for the main subhalo in a FOF halo)}. The Press-Schechter \citep{PS74} and Sheth-Tormen functions \citep{ST99} are over-plotted.}
  \label{massf}
\end{figure}

We now compare the halo mass distribution with predictions of analytical models. Let $\Delta n$ be the \MScolor{number density} of objects with mass (FOF mass found for our groups \MScolor{or mass of main subhalo in our group}) within a logarithmic bin, and $\Delta \log M$ be $\log M_2/M_1$ where $M_2/M_1$ is the ratio of upper and lower values for each mass bin (constant).  In Fig.~\ref{massf}, $\frac{\Delta n}{\Delta \log M}$ is plotted against the median value in each mass bin along with Poisson error bars \MScolor{(error bars are omitted for subhalo data for easy recognition)}. The analytical Press-Schechter \citep*{PS74} and Sheth-Tormen models \citep*{ST99} are plotted with continuous and dotted lines, respectively. The Press-Schechter function underestimates the number of haloes in the high mass range, whereas the Sheth-Tormen function gives a good fit to the \MScolor{simulated FOF data} over a broad range of masses including the \MScolor{high-mass} end. \MScolor{Since subhaloes contain only part of mass of its host FOF haloes} \MScolor{and therefore some of the haloes that belonged to a particular mass bin is shifted to neighboring lower mass bin}, \MScolor{number density of subhaloes in a given mass bin is systematically smaller than FOF haloes.} Overall, we find that these analytical fitting formulae reproduce not only the statistics of dark matter haloes with mass $10^7 \,h^{-1}{\rm M}_{\odot}$ as previously known \citep{Ish13} at $z = 0$, but they also describe less massive haloes obtained from our simulations, such as those as small as $10^3 \,h^{-1}{\rm M}_{\odot}$ at $z = 15$.

\subsection{Formation time}
\label{sec_zform}

\begin{figure}
  \includegraphics[width=7.5cm]{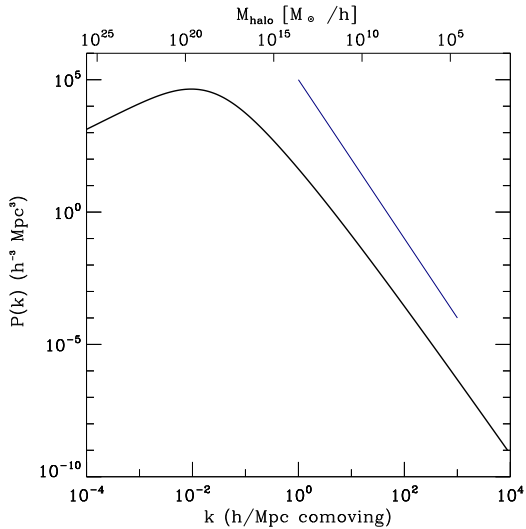}  
  \caption{The linearly extrapolated power spectrum of dark matter at $z = 0$ for the cosmological parameters we adopt. The critical slope of $k^{-3}$ is shown with dark blue line.}
  \label{PowerS}
\end{figure}

The linear power spectrum of density perturbations at $z = 0$ for the cosmological parameters we adopt in our work is plotted in Fig.~\ref{PowerS}, where the lines show the critical slope of $-3$. In this Section, we express the slope of the power spectrum by $p$. Therefore, $P(k) \propto k^p$, where the slope $p$ changes as a function of wave-number $k$. \MScolor{For the linear power spectrum shown in Fig.~\ref{PowerS}, $p = -2.37, -2.62, -2.74, -2.80$ at $k = 1, 10, 100, 1000 h {\rm \; Mpc^{-1}}$.}

The power spectrum increases at very low wave numbers, then turns over and starts to decrease again, approaching power law with $p \simeq -3$ at high wave numbers. For the mass scales extensively studied by earlier works, ${M}_{\rm halo} \sim 10^{12} \,h^{-1}{ \rm M}_{\odot}$, the slope is substantially shallower compared to a mass scale of $M_{\rm halo} \sim 10^{6} \,h^{-1}{ \rm M}_{\odot}$ where the slope is close to the critical value of $p = -3$.  It is critical in the sense that the non-dimensional power, \begin{equation}
  \Delta^2(k) = \frac{k^3 P(k)}{2 \pi^2} 
\end{equation}
becomes independent of wave number $k$. $\Delta^2(k)$ represents the
amount of perturbations in a logarithmic bin in $k$. For $p > -3$, $\Delta^2(k)$ is
a monotonically increasing function of $k$. Therefore, given $p > -3$, for a fixed time, there is more power in small scales (large $k$) than in large scales (small $k$). Thus, small gravitational structures form first in such regimes (explicitly shown in Fig. 1 in \citealt{Har06}). 
In contrast, for $p = -3$, $\Delta^2(k)$ is independent of wave number $k$, thus the
dark matter structures of various mass collapse simultaneously and in a non-hierarchical fashion. We now show
this explicitly using our high-resolution simulation.

\begin{figure}
  \includegraphics[width=8cm]{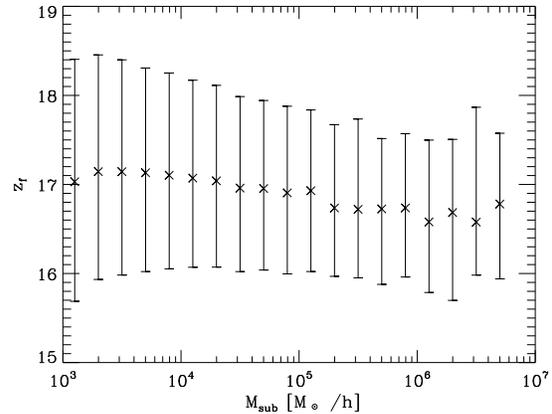} 
  \caption{Formation time of dark matter haloes as a function of their mass. Median, 20 \%, and 80 \% values in each mass bin are plotted. Haloes that differ in mass by $\sim$ four orders of magnitude form simultaneously. This is quite different from the formation time for more massive systems at $z = 0$. Compare with Fig.~\ref{Mil-tf}.}
\label{tform}
\end{figure}

Adopting the methods described in Section~\ref{mtree}, we identified the formation time of each FOF halo. In Fig.~\ref{tform}, we divide dark matter haloes in various mass bins and plot the median, and 20\%, 80\% values of formation time in each mass bin. We find that in the mass range we are interested in, massive dark matter haloes and small haloes form simultaneously (this is only true statistically, because there is a dispersion of $\sim 10\,{\rm Myr}$ in the formation time in a given mass bin). This is quite different from the more massive regime, where less massive haloes form first (\citealt{Har06}, also see results of the `Large' run in Fig.~\ref{Mil-tf}). This implies that the formation processes of dark matter mini-haloes found at high redshift in a simulation box of $1\,h^{-1}{\rm Mpc}$ differ substantially from those found at much higher halo masses in larger boxes.

\subsection{Halo shape}

\label{sec_shape}
Often the high density regions found in collisionless $N$-body simulations are not exactly spherically symmetric, as opposed to the assumption which is made to derive Press-Schechter mass function for example. Modelling dark matter haloes as ellipsoids and measuring their three axes provides a way to quantify the shape of these haloes and how much it differs from the spherical collapse model. Moreover, there are several different kinds of approaches made observationally to estimate halo shapes in the local universe: Firstly, there are ways to measure dark matter distribution by weak lensing using X-ray clusters \citep{Ogu10}. Secondly, there are also attempts to determine dark matter halo shapes for galaxies statistically using surveys \citep*{Hoek04}. Lastly, there are also efforts made to derive the gravitational potential of our Galaxy by using the kinematics of tidal tails of the Sagittarius dwarf spheroidal galaxy \citep*{Law09}. The less massive haloes found in our study are influenced by a much shorter wavelength portion of the $\Lambda$CDM power spectrum than the haloes examined in these observational studies, and it is therefore interesting to see whether there is any systematic difference in the shapes of these mini-haloes.

We define a second moment tensor of the halo shape as:
\begin{equation}
  I_{ij} = \sum_n x_i x_j,
\label{moment}
\end{equation}
\noindent
where $x_i$ ($i=1$, $2$, $3$ corresponding to $x$, $y$, and $z$ coordinates) is the particle position with respect to the centre of the halo, which we define as the position of the particle with the lowest potential energy. The same form of the shape matrix was adopted in the earlier studies of \citet{Jan01} and \citet*{Macc08}. We note that sometimes $I_{ij}$ is defined as $\sum_n x_i x_j/r'^2$, i.e.~with a $1/r'^2$ normalization factor (\citealt{Moo04}, \citealt{All06}). But we here adopt the former, more widely used formulation.

The sum is taken over \MScolor{all the particles that belong to the main subhalo in a FOF halo}. After evaluating the inertia tensor, we compute its eigenvalues $I_1$, $I_2$, and $I_3$, which are the three principle moments of inertia of the halo, and which satisfy the relationship $I_1 \ge I_2 \ge I_3$. The lengths of the axes $a$, $b$, and $c$ associated with the principle moments of inertia are given by 
\begin{eqnarray}
a & = & \sqrt{(5\ I_1)/N_p}, \\
b & = & \sqrt{(5\ I_2)/N_p}, \\
c & = & \sqrt{(5\ I_3)/N_p}, 
\end{eqnarray}
where $N_p$ is the number of particles summed up in equation~(\ref{moment}).
The sphericity of the halo is then defined as 
\begin{equation}
s = c/a . 
\label{defs}
\end{equation}
From the above formulation, it should be clear that a spherically symmetric halo has $s = 1$. The triaxiality of the halo is defined as 
\begin{equation}
T = (a^2 - b^2)/(a^2 - c^2) . 
\label{deft}
\end{equation}
By definition, an oblate halo ($a = b  >  c$) has $T = 0$, and a prolate halo ($a > b = c$) has $T = 1$.

\begin{figure*}
  \begin{center}
    \begin{tabular}{cc}
      \includegraphics[width=8.0cm]{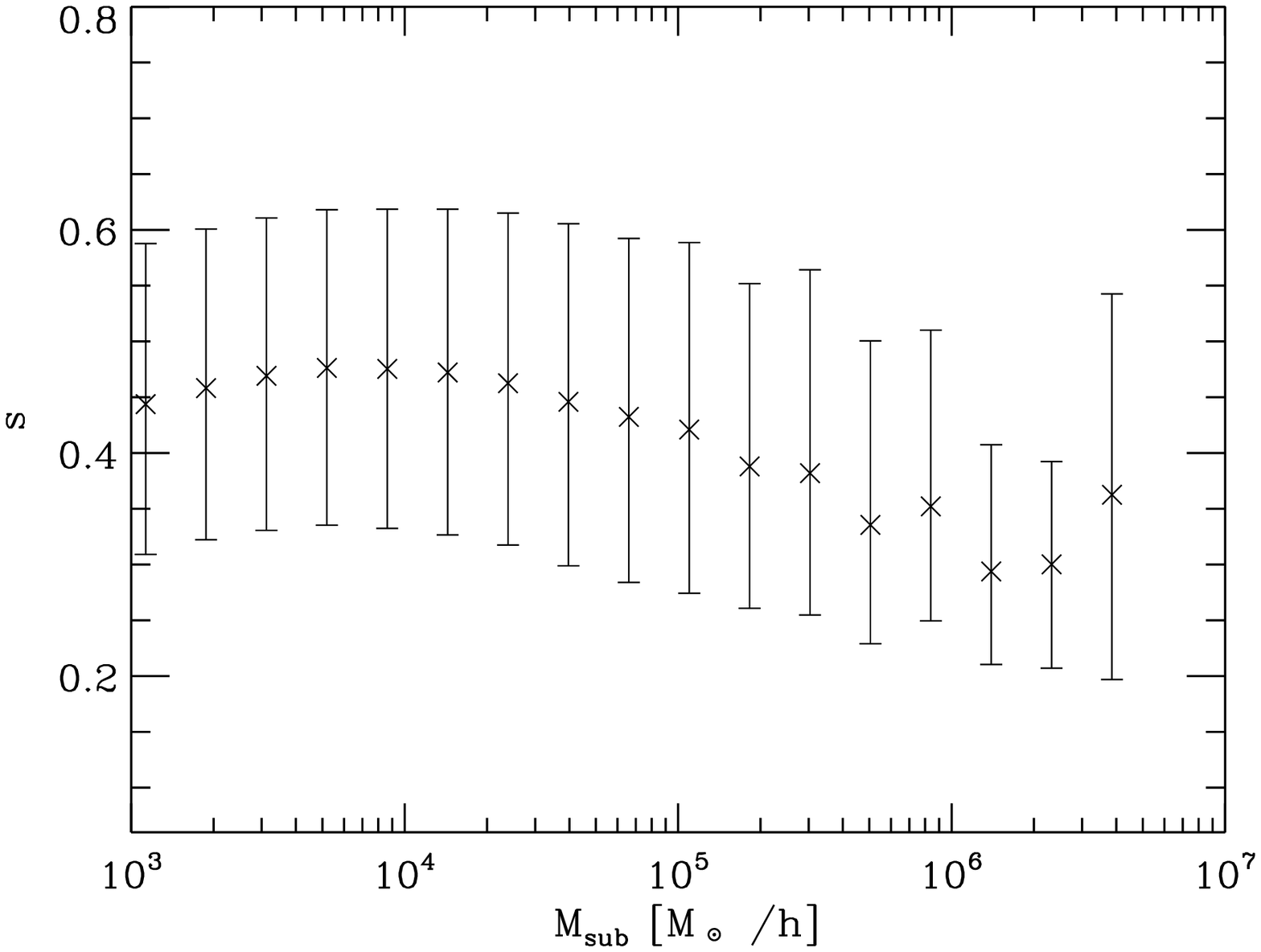}  &   
      \includegraphics[width=8.0cm]{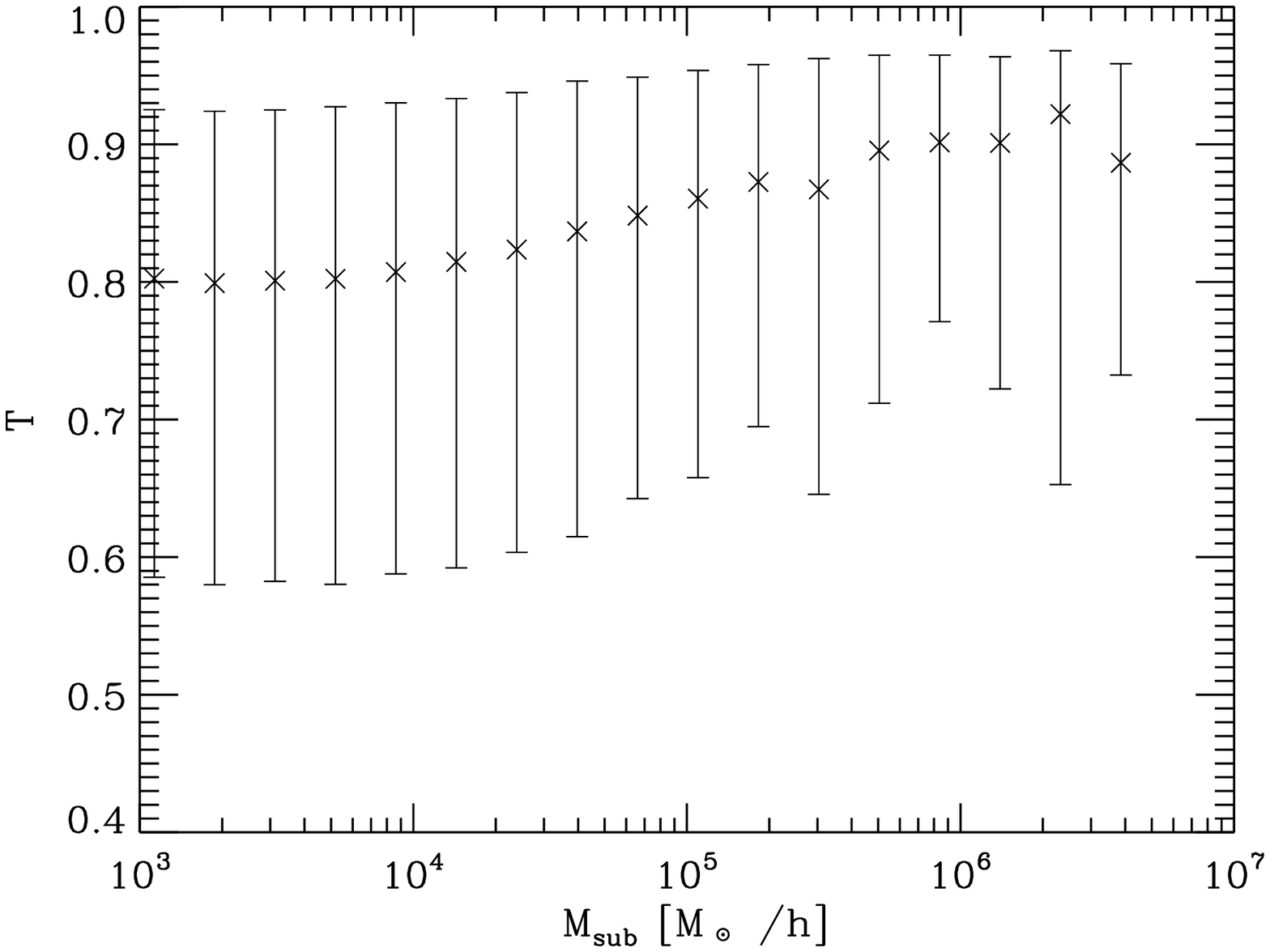} \\   
    \end{tabular}
    \caption{Median of sphericity $s$ \& triaxiality $T$ in different mass bins for dark matter haloes at $z = 15$. The sphericity and triaxiality are defined by equations~(\ref{defs}) and (\ref{deft}), respectively.}
    \label{stmed}
  \end{center}
\end{figure*}

\begin{figure*}
  \begin{center}
    \begin{tabular}{cc}
      \includegraphics[width=8cm]{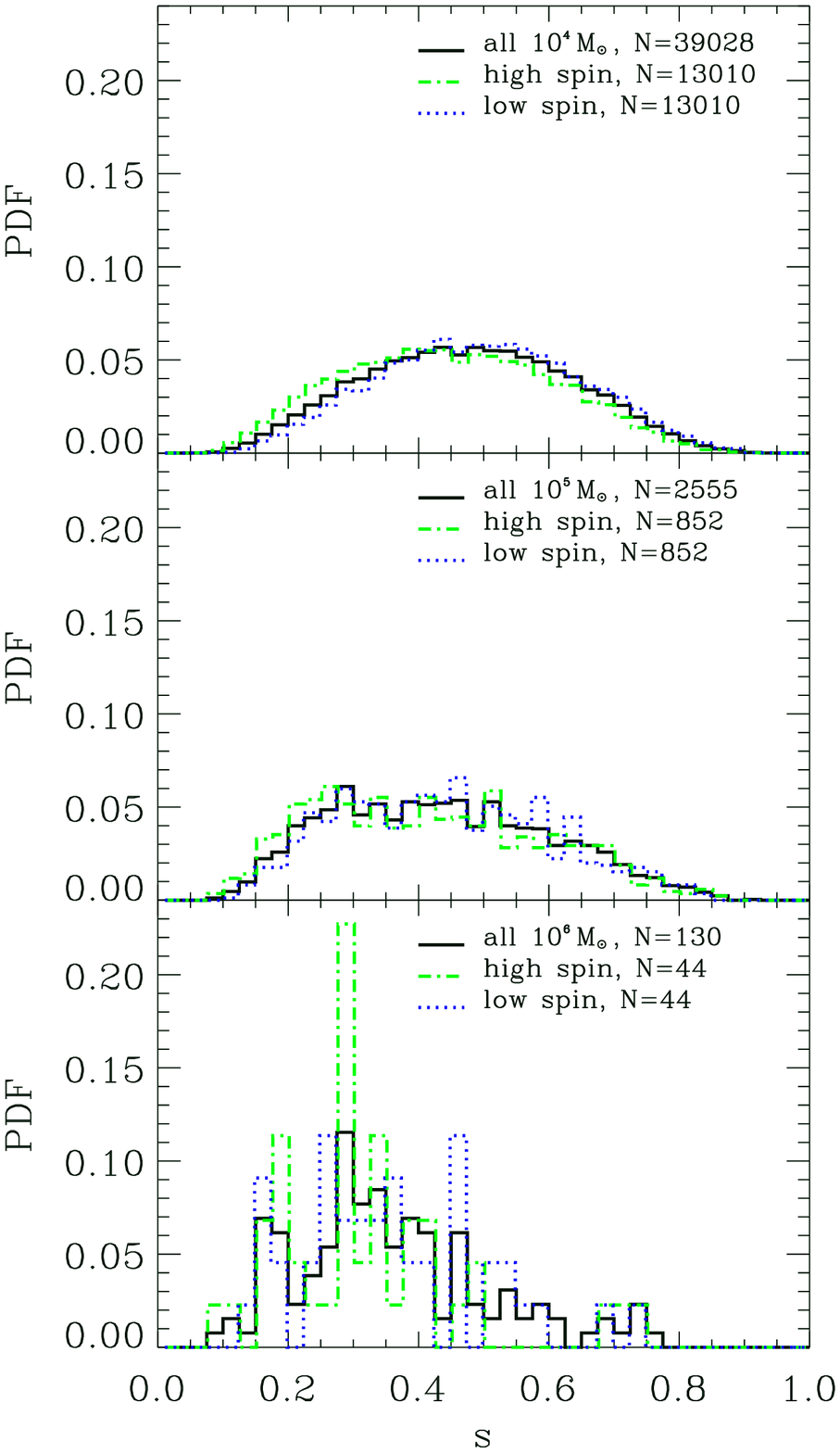}  &   
      \includegraphics[width=8cm]{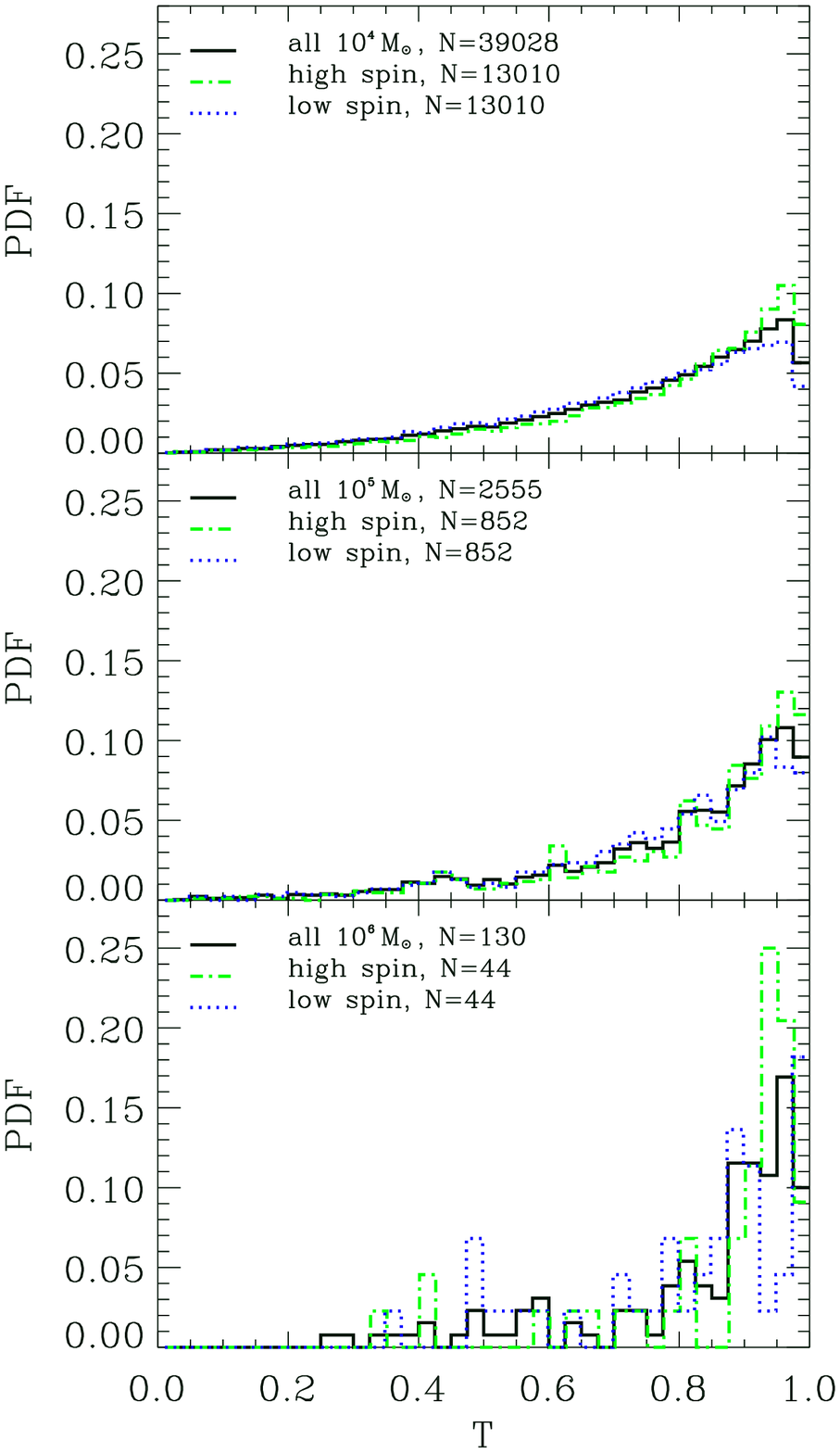} \\ 
    \end{tabular}
    \caption{Sphericity $s$ \& triaxiality $T$ for haloes with mass $10^{4\pm 0.2} \, h^{-1} {\rm M}_{\odot}$, $10^{5\pm 0.2}\, h^{-1} {\rm M}_{\odot}$, and $10^{6\pm 0.2} \, h^{-1} {\rm M}_{\odot}$ at $z = 15$. Different lines represent distributions for different groups in each mass range. (Blue dotted lines correspond to halos with $\lambda < \lambda_{1/3}$, dash-dot lines correspond to halos with $\lambda > \lambda_{2/3}$, and black solid lines correspond to all \MScolor{sub}haloes in that mass range.)}
\label{sandspin}
  \end{center}

\end{figure*}

Black solid lines in the different rows in Fig.~\ref{sandspin} show the distribution of sphericity $s$ for different mass ranges (namely haloes with mass $10^{4\pm 0.2}\,h^{-1} {\rm M}_{\odot}$, $10^{5\pm 0.2}\,h^{-1} {\rm M}_{\odot}$, and $10^{6\pm 0.2}\,h^{-1} {\rm M}_{\odot}$). We find that the distribution of $s$ based on all haloes has a peak at \MScolor{$0.4-0.45$}, which is smaller than the value typically found for more massive haloes at lower redshifts. For example, \citet{All06} and \citet{Macc08} investigated haloes with masses $\sim 10^{12}\,h^{-1} {\rm M}_{\odot}$ to $10^{15}\,h^{-1} {\rm M}_{\odot}$ and found almost no haloes with $s < 0.4$. This could be due to more frequent mergers at $z = 15$ compared to $z = 0$.

If we compare the distribution function of $s$ and $T$, $10^5\,{\rm M}_{\odot}$ haloes are more likely than $10^4\, {\rm M}_{\odot}$ haloes to have lower values of sphericity $s$ and higher values of triaxiality $T$. This tendency of more massive haloes having small $s$ and larger $T$, that has been observed elsewhere (\citealt{Macc08}, \citealt{All06}), is even clearer in Fig.~\ref{stmed}, in which we show median values of $s$ and $T$ as a function of halo mass. Comparing Fig.~\ref{stmed} with Fig.~\ref{Mil-med}, we find that haloes found at $z = 0$ are systematically more spherical and only moderately more prolate compared to haloes at $z = 15$.

Fig.~\ref{sandspin} also shows that for all the three different mass ranges studied here, a large fraction of halos are prolate ($T\sim 1$). This is similar as found in previous studies that focused on haloes forming in different regimes (e.g.~more massive haloes, \citealt*{Dub91}, \citealt{War92}). This is most likely because dark matter haloes of various mass scales form from filamentary structures. However, we find a significant disagreement with \citet{Souz13}, who worked on similar mass ranges and practically the same redshift, but with hydrodynamical simulations. Their dark matter haloes with mass $\sim 10^5 \, {\rm M}_{\odot}$ at $11 < z < 16$ have $s \sim 0.3$ and more than 90\% have $T \lesssim 0.4$, whereas the dark matter halos with mass $\sim 10^5 \,h^{-1}{\rm M}_{\odot}$ at $z = 15$ in our simulations have medians $s \sim \MScolor{0.4}$ and  $T \sim \MScolor{0.85}$. We note that \citet{Souz13} adopt a definition of $I_{ij}$ equivalent to ours. The difference in shape could be due either to the inclusion of gas cooling or to a very different mass resolution. We resolve \MScolor{substructures inside} dark matter haloes \MScolor{with mass} $\sim$ $10^5\,h^{-1} {\rm M}_{\odot}$ with $\sim 10^4$ particles, whereas \citet{Souz13} resolve haloes with the same mass with only $\sim 200$ particles. We speculate that the inclusion of gas makes a bigger difference than the mass resolution, however, since our convergence study shows that even with a relatively small particle number of 200 particles or so, the errors in the axial ratios ($b/a$, $c/a$) remain within 10\% (see Table~\ref{tab:abc}).

Histograms with different color in Fig.~\ref{sandspin} represent halo shape distributions for haloes with different values of the spin parameter. We sorted dark matter haloes in each mass range into percentiles according to their spin. If we denote the 33\% and 67\% percentile values of the spin parameter in each mass range as $\lambda_{1/3}$ and $\lambda_{2/3}$, we have plotted the probability distribution of halo shapes for three different kinds of halo selections: all the haloes in a certain mass range, only haloes with $\lambda > \lambda_{2/3}$, and only haloes with $\lambda < \lambda_{1/3}$. We found that, in all the mass ranges we looked at, dark matter haloes with high spin are less spherical and more highly prolate, a trend already confirmed in lower redshift dark matter haloes by previous works \citep{Bet07}. \citet{Dav10}, who used different methods to identify haloes and estimate their shapes, also found similar trends for high-redshift dark matter haloes with relatively low resolution.  This widely observed correlation could occur because haloes that experienced strong gravitational forces during their formation are likely to have high spin and an aspherical shape.  

In short, using the highest-resolution simulation to date for resolving a large sample of \MScolor{substructure inside} dark matter mini-haloes of mass $> 10^3\,{\rm M}_{\odot}$, we have demonstrated that dark matter mini-haloes we find in our simulations at $z \sim 15$ have qualitatively many similarities with dark matter haloes found at $z <\sim 6$ regarding their shape.
 
\subsection{Correlation between formation time and virial ratio / formation time and halo shape}
\label{sec_trts}
Many authors have studied the effect of the mass accretion history on the concentration and density profile of dark matter haloes (\citealt{Bul01}, \citealt{Wec02}, \citealt{Tas04}). Much less work has been done on the relationship between formation time and other halo properties.  For dark matter haloes found in mass scales capable of hosting Pop III star formation, we show for the first time that there is a direct connection between the shape of a halo, its dynamical state, or \MScolor{its spin} and its formation time. This could provide interesting clues on the formation processes of dark matter haloes.

\begin{figure}
  \includegraphics[width=8cm]{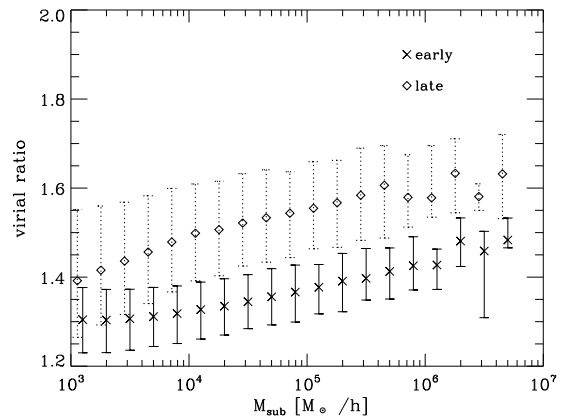}
  \caption{Relation between formation time and virial ratios. We grouped
 the haloes in each mass bin into three groups. Early one-third (old haloes), middle
 one-third (intermediate haloes), and late one-third (young haloes). We plot the median values for early (crosses)
 and late (diamonds) thirds. The error bars indicate 20th and 80th percentile values for each group. (Error bars are plotted in solid lines for haloes forming earlier, and in dotted lines for haloes forming later. For easy reference, haloes forming later are slightly offset \MScolor{in mass}.) We found that haloes that formed later (young haloes) have higher virial ratios on average due to lack of time for relaxation.}
\label{t-r}
\end{figure}

\begin{figure}
  \includegraphics[width=8cm]{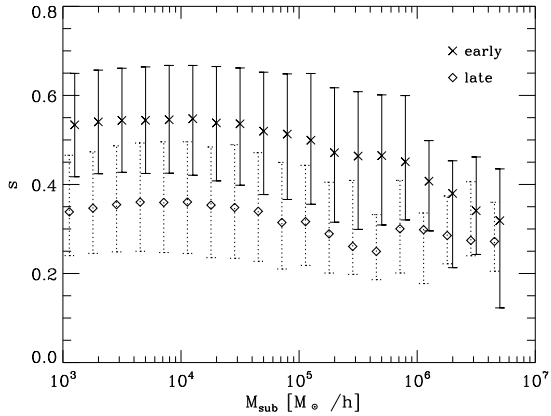}
  \caption{Relation between formation time and sphericity. We grouped
 the haloes in each mass bin into three groups. We plot the median values with error bars for early (crosses) and late (diamonds) thirds as in Fig.~\ref{t-r}. We found that haloes that formed later (young haloes) are less spherical due to recent mass accumulation.}
\label{t-s}
\end{figure}

\begin{figure}
  \includegraphics[width=8cm]{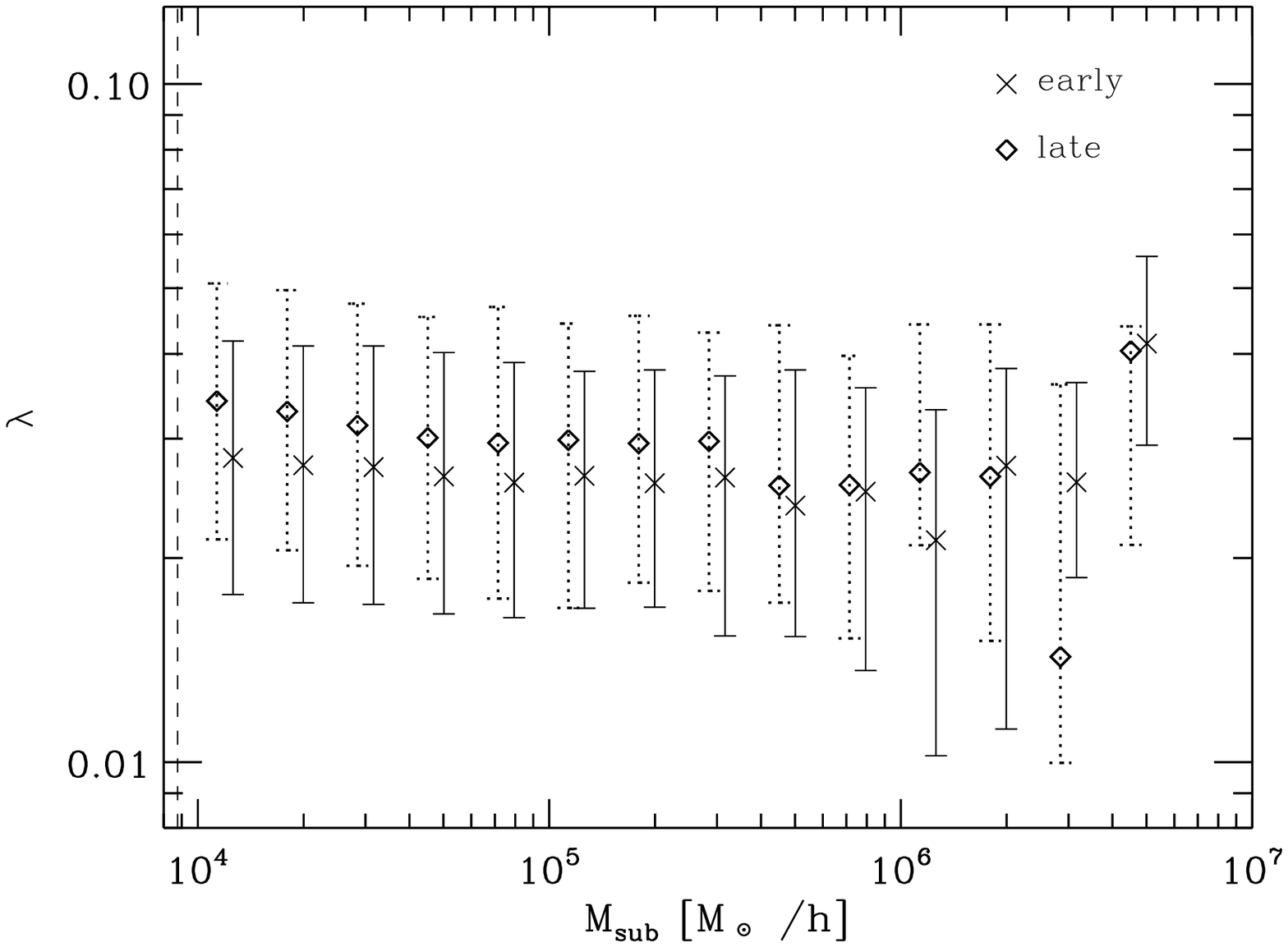}
  \caption{\MScolor{Relation between formation time and spin parameter. We grouped
 the haloes in each mass bin into three groups. We plot the median values with error bars for early (crosses) and late (diamonds) thirds as in Fig.~\ref{t-r}. We found that haloes that formed later (young haloes) have values of spin parameter due to recent mass accumulation.} \MScolor{The dashed vertical lines represent lower mass limit of 1000 particles.}}
\label{t-l}
\end{figure}

We sorted the dark matter haloes in different mass bins into three groups according to their formation time. In Fig.~\ref{t-r}, we show that the haloes that formed later (young haloes) have higher values of the virial ratio, \MScolor{while in Figures \ref{t-s} and \ref{t-l}, we show that haloes that formed later (young haloes) have smaller sphericities and larger spin parameters}. This is because haloes of a given mass that formed later accumulated their mass recently, and thus had no time for relaxation.

\MScolor{To clarify this point, let us make comparison of relevant timescales. The Hubble time, $t_{\rm H} = a/\dot{a}$ is $\sim$ 300 Myr at $z = 15$. If we estimate the relaxation time $t_{\rm relax}$ of dark matter haloes by $t_{\rm relax} = \frac{1}{\sqrt{G \rho}} = \frac{1}{\sqrt{G 200\rho_{\rm crit}(z = 15) }}$, $t_{\rm relax}$ $\sim$ 60 Myr at $z = 15$ ($ < t_{\rm H}$, as expected). The old haloes analyzed here typically formed at around $z = 18 - 20$ whereas the young haloes formed at $z < 16$. Thus the former typically had more than $\Delta t (z = 18, z = 15) \sim$ 80 Myr ($ > t_{\rm relax}$)} \MScolor{while} \MScolor{the latter had less than $\Delta t (z = 16, z = 15) \sim$ 45 Myr ($ < t_{\rm relax}$) since the time of formation. This could account for the differences in physical properties we observe when we stop our simulation at $z = 15$.}

In fact, \citet{Hetz06} have demonstrated that major mergers increase the value of the spin parameter and the virial ratio. \citet{All06} have also found that haloes forming earlier (old haloes) are more spherical. They have also shown that the dependence on formation time is weaker for higher mass haloes, at least in their simulation box. The different methods adopted to calculate halo shape and formation time preclude a direct comparison, but the high-mass end of Fig.~\ref{t-s} clearly shows similar trends.

Our results confirm that the formation epoch of dark matter haloes influences global parameters such as shape, virial ratio, \MScolor{and spin}. Each dark matter halo has a different evolution history. Furthermore, even the dark matter haloes that have similar mass accretion histories could have accreted their mass from different spatial locations or through different channels. \MScolor{A single global parameter such as formation time is not enough to account for this,} and the scatter in Fig.~\ref{t-r} and Fig.~\ref{t-s} could be due to variations in the processes that let haloes accumulate their mass.

\subsection{Correlation function}
\label{sec_xi}
The two-point correlation function calculated from $N$-body simulations is a useful quantity, since it can be compared directly with galaxy clustering data \citep{Spr05}. We perform a similar analysis to clarify the properties of the mini-haloes we find at $z = 15$. \citet*{Gao05} have demonstrated that haloes that assembled earlier are more strongly clustered, casting doubt on an assumption made in excursion-set theory \citep{Bond91}, namely that halo properties \MScolor{depend}3 only on mass and are independent of environment. 

Here we study the clustering of haloes with different spins.  In order to quantify the clustering between dark matter mini-haloes, we make two kinds of catalogs of dark matter haloes: a simulated halo catalog and a random catalog. The simulated halo catalog is a list of positions of haloes  obtained from our simulation. The random catalog is a list of random points distributed in a box that is equal in size to our simulation box.  We measure the two-point correlation function $\xi(r)$ following \citet*{Ham93}, \begin{equation}
  \xi(r) = \frac{DD(r)RR(r)}{RD(r)^2} - 1 ,
\end{equation}
where $DD(r)$, $RD(r)$, $RR(r)$ stand for, respectively, the number of pairs with separation $r$ in the simulated halo catalog (halo-halo pairs), the simulated halo catalog and random catalog (halo-random point pairs), and just the random catalog (random point-random point pairs). The statistical errors are estimated through $RR(r) \sqrt{DD(r)} /  RD(r)^2$.

\begin{figure}
  \includegraphics[width=8cm]{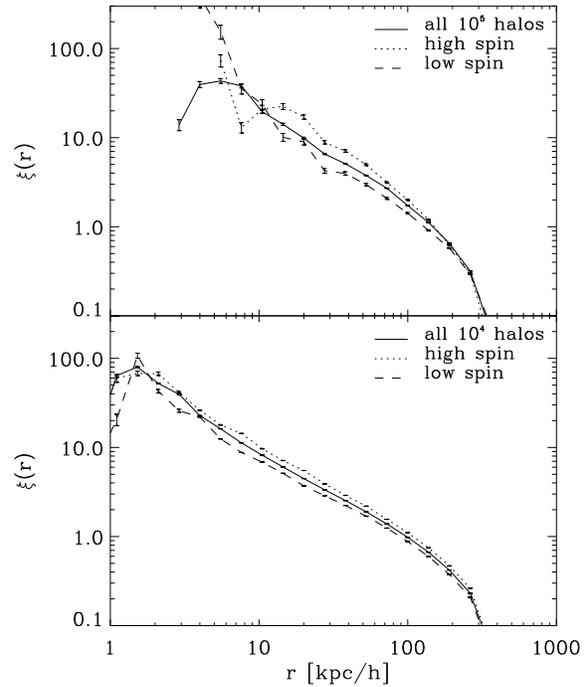}
  \caption{Two-point correlation function for dark matter haloes in the mass range $10^{5\pm 0.2} \, h^{-1} {\rm M}_{\odot}$ and $10^{4\pm 0.2}\, h^{-1} {\rm M}_{\odot}$. Distance is in units of comoving $h^{-1}{\rm kpc}$. We have reproduced the trend seen in previous works that high spin haloes are more highly correlated than low spin haloes. }
\label{xi}
\end{figure}

The two-point correlation functions for haloes in the mass bin $10^{5\pm 0.2} \, \rm M_{\odot} $ and $10^{4\pm 0.2} \, \rm M_{\odot} $ are depicted in Fig.~\ref{xi} as a function of separation in units of comoving $h^{-1}{\rm kpc}$. As in Section~\ref{sec_shape}, we sorted dark matter haloes in both mass ranges into thirds according to their spin. We found that haloes with high spin are more clustered, which is well-known in large-scale simulations (\citealt{Bet07}, \citealt*{GaoL07a}), but was only investigated with relatively low resolution in much smaller scales (\citealt{Dav09}). This suggests that haloes that form in a clustered environment are more likely to experience tidal forces from neighboring overdense regions, and therefore tend to have larger spin parameters.
 
\section{Conclusions}

In this paper, we have performed a high-resolution numerical simulation of structure formation up to the high redshift $z = 15$, inside a simulation box of side-length $1\,h^{-1}{\rm Mpc}$. Exploiting our good statistics and resolution, we made high precision measurements for a variety of global physical parameters of halos and demonstrated correlations between some of these properties.  We clarified the characteristics of dark matter mini-haloes by comparing the results of our high-resolution simulation with works found in the literature on more massive haloes and at lower redshift ($z < 6$), and with results from our own moderate-resolution complementary simulation within a cubic region of $100\,h^{-1}{\rm Mpc}$ on a side that we evolved until $z = 0$.  Our main findings are:

\begin{enumerate} 
\item Dark matter haloes found in our simulations have a distribution of spin parameters that is well fitted by a log-normal function around $\bar{\lambda} = \MScolor{0.0262}$, with dispersion $\sigma_{\lambda} = \MScolor{0.495}$. This value for $\bar\lambda$ is similar to the value $\sim 0.03$ obtained by studying more massive objects at $z \sim 0$ (e.g. galaxy clusters). The dimensionless spin parameter $\lambda$ is somewhat sensitive to resolution, and hence our measurement of this quantity is by far the most accurate thus far for the mass scales investigated in this work.

\item We have shown explicitly from output of our simulations and merger trees constructed from them that the formation time of dark matter haloes, defined as the time at which the most massive substructure in a FOF halo reaches half of its final mass, only weakly depends on halo mass over about four orders of magnitude (e.g.~mass scales of $\sim 10^{3-7} \, \rm M_{\odot}$). At larger mass scales, structures are known to form in a hierarchical fashion (formation redshift is a monotonically decreasing function of halo mass).  But on the mass scales investigated here, this at least partially breaks down. The weak dependence of the formation time on mass is a result of the slope of the power spectrum of density perturbations at this scale, where it becomes close to the critical value of $-3$ for which all mass perturbations 
are expected to collapse simultaneously. The scatter in formation time is $\sim$ 10 Myr.

\item The shapes of haloes are much less spherical and more highly prolate than haloes found at $z \sim 0$. The most frequent value of the sphericity parameter lies between \MScolor{0.4 and 0.45}. This could be due to more frequent mergers at high redshift. The majority of halos has triaxiality parameter $> 0.9$. The more massive haloes are more likely to be \MScolor{slightly} less spherical and \MScolor{slightly} more filamentary (prolate). 

\item We have also investigated the relationship between formation time and halo properties such as shape, virial ratios, \MScolor{and spin} which was not studied previously for the scales examined here. On average, haloes that formed more recently (young haloes) have higher values of virial ratio, are less spherical, \MScolor{and have higher values of spin} when observed at $z = 15$. \MScolor{This can be understood because the time passed since formation of these young haloes until the end of the simulation is less than relaxation time, $t_{\rm relax} \sim$ 60 Myr, of these haloes.}

\item Although not expected from excursion set theory, we find that haloes with high spin are more strongly clustered, where clustering is quantified by the two-point correlation function of the positions of different dark matter halos. This could mean that haloes born in clustered environments experience stronger tidal torques during their formation.  

\end{enumerate}

In this study, we have investigated broad aspects of dark matter haloes at high-redshift and found many similarities with their low-redshift counterparts. Our findings could have important implications for the baryonic component of dark matter haloes found at high-redshift.  Our results on the correlation between formation time and virial ratio/shape imply that the gas component inside each dark matter halo could evolve differently depending on accretion history and the actual dynamics of accretion.

\MScolor{Potential caveats of our work include starting redshift of our simulations and exclusion of surface terms in estimating the virial ratios. We will briefly discuss them in the following paragraphs.}

\MScolor{\citet{Reed07} shows that an initial redshift of 139 should be safe for studying objects at $z \simeq$ 7-15. Since suppression of high sigma density peaks that result from use of first order perturbation theory and low starting redshift is known to be stronger at high redshift \citep{Croc06}, it is possible that we are missing some of the dark matter haloes at the high mass end, especially when studying properties such as spin parameter of them at $z \sim 25$. Due to this suppression effect, we could be under-estimating the two-point correlation function and formation time of dark matter haloes as well.}

\MScolor{It is pointed out in recent literature (\citealt{Bal06}, \citealt{Dav11}) that the surface terms in virial equations are in general not negligible. Once these terms are taken into account, dark matter haloes tends to have less excess kinetic energy. However, \citet{Dav11} show that even after correcting for the surface terms, dark matter haloes have virial ratios that are greater than one and that increase with increasing redshift. Therefore, although we would expect a systematic decrease in the absolute values of virial ratios once we include the surface terms, our main results regarding the dependence of the virial ratio on halo mass and formation redshift should be sound.}

In a future study, it would be interesting to directly follow the star formation taking place in different dark matter mini-haloes by means of hydrodynamical simulations, and to clarify how the global dark matter properties investigated here are connected to the properties of the resulting Pop III stars.

\section*{Acknowledgements}

M.S. acknowledges computing time from H\"{o}chstleistungs Rechenzentrum Nord (HLRN) under project hhp00022 and financial support from the Heidelberg Graduate School of Fundamental Physics (HGSFP).   
R.S.K. acknowledges support from the European Research Council under the European Communities Seventh Framework Programme (FP7/2007-2013) via the ERC Advanced Grant “STARLIGHT: Formation of the First Stars” (project number 339177). M.S. and R.S.K. thank for funding from the Baden-W\"{u}rttemberg Foundation in the program Internationale Spitzenforschung (contract research grant P-LS-SPII/18).
M.S. wishes to thank Matthias Bartelmann for clarifying some of the basic concepts in structure formation. 
\bibliographystyle{mn2enew}
\bibliography{ref}

\appendix

\section{Comparison of structure formation simulation at large and small scales.}
\label{sec_app}

As a validity check for the data analysis method we adopt, and to make a qualitative comparison between structure formation occurring at different scales, we performed a numerical experiment to follow the evolution of gravitational structures at larger scales with relatively small particle numbers.  The basic simulation parameters of this comparison simulation and the main simulation are listed in Table~\ref{Mil}, where $L_{\rm box}$ is the size of the simulation box, $N_{\rm par}$ is the number of dark matter particles included in the simulation, and $z_{\rm fin}$ is the redshift at which we terminate each of the simulations. Since the size of the simulation box differs by two orders of magnitude in the two runs, the mass scales of the gravitational structures found in each of the two runs differ by $\sim$ $\rm 10^6$.
  
\begin{table}
\caption{Simulation parameters}
\begin{center}
\begin{tabular}{lcccc}\hline\hline
      &          &    $L_{\rm box}$             &       &   $\epsilon$ \\
Run   &  $N_{\rm par}$    &   ($h^{-1}$ Mpc)  &  $z_{\rm fin}$  &   ($h^{-1}$ kpc)  \\ \hline

Large &  $512^3$  &    100     &  0     &   4.0    \\

Small &  $2048^3$ &      1     &  15  &    0.01  \\  \hline 

\end{tabular}
\end{center}
\label{Mil}
\end{table}

\begin{figure*}
  \begin{center}
    \begin{tabular}{cc}
      \includegraphics[width=8.0cm]{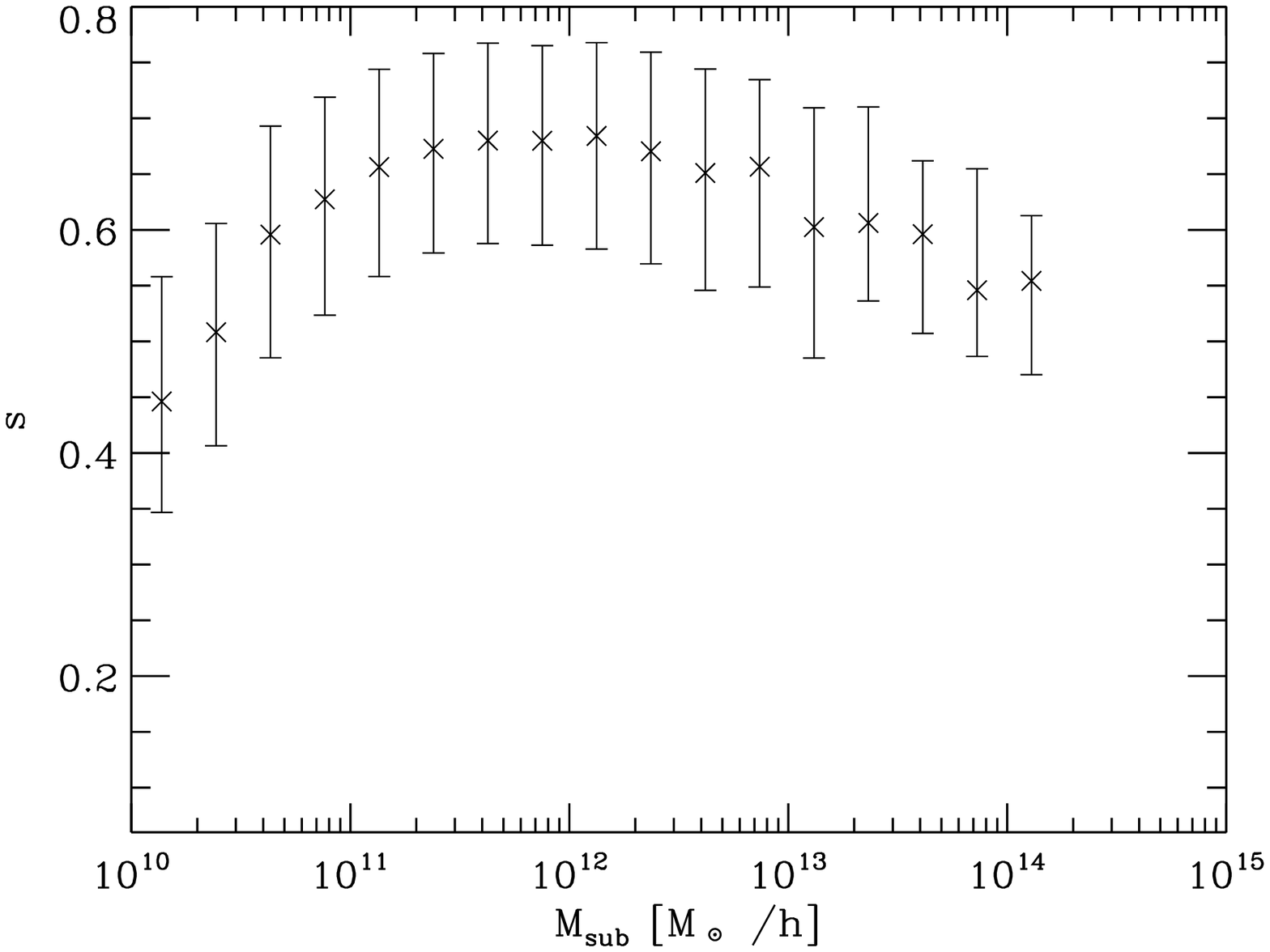}  &   
      \includegraphics[width=8.0cm]{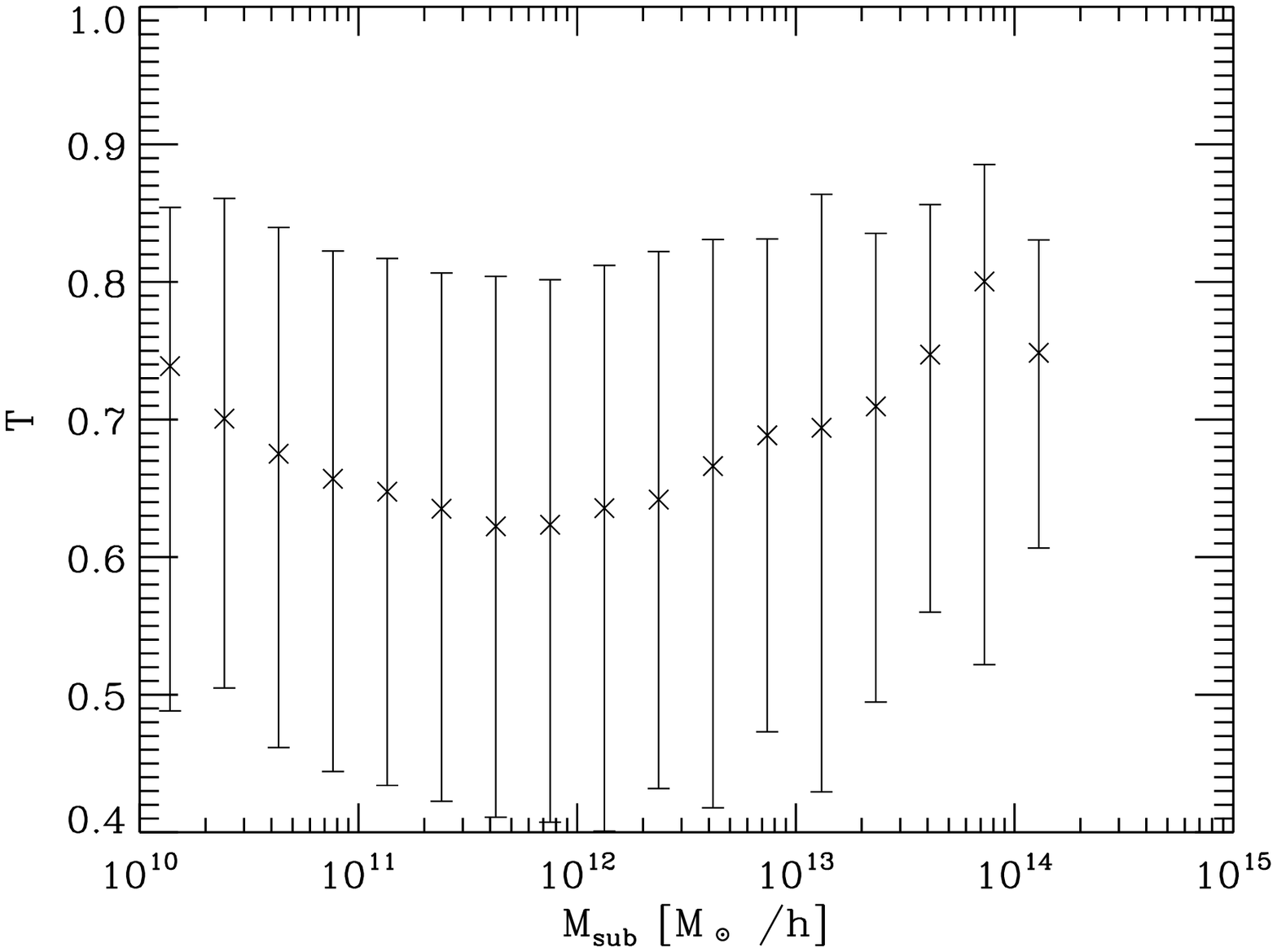} \\   
    \end{tabular}
    \caption{Median of sphericity $s$ \& triaxiality $T$ in different mass bins for dark matter haloes at $z = 0$. The sphericity and triaxiality are defined by equations~(\ref{defs}) and (\ref{deft}), respectively.}
    \label{Mil-med}
  \end{center}
\end{figure*}

\begin{figure}
 \includegraphics[width=8cm]{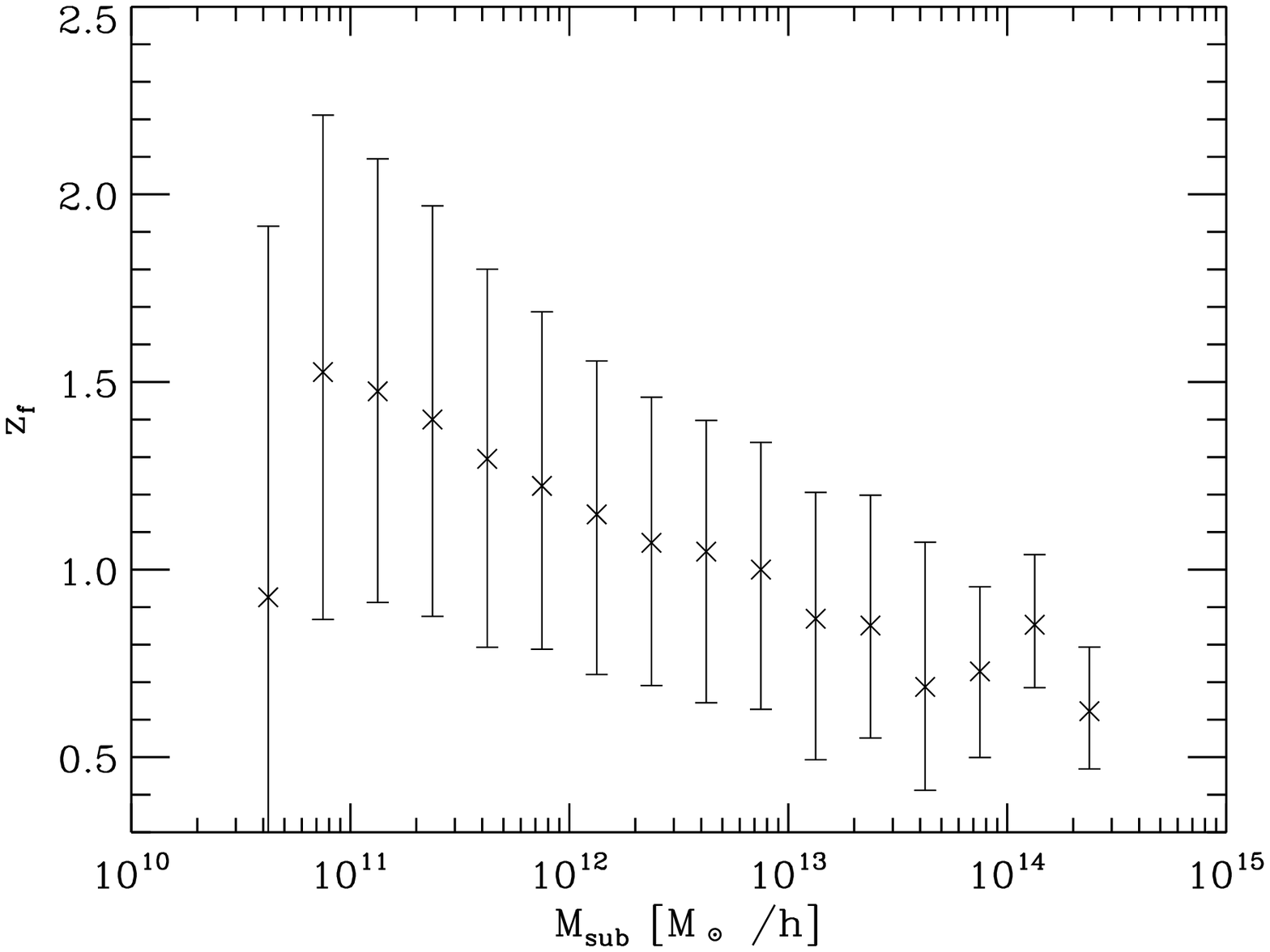}
 \caption{Formation time of dark matter haloes as a function of their mass. Median, 20 \%, and 80 \% values in each mass bin are plotted. 
More massive dark matter haloes form at smaller redshift.}
 \label{Mil-tf}
\end{figure}

\begin{figure}
  \includegraphics[width=8cm]{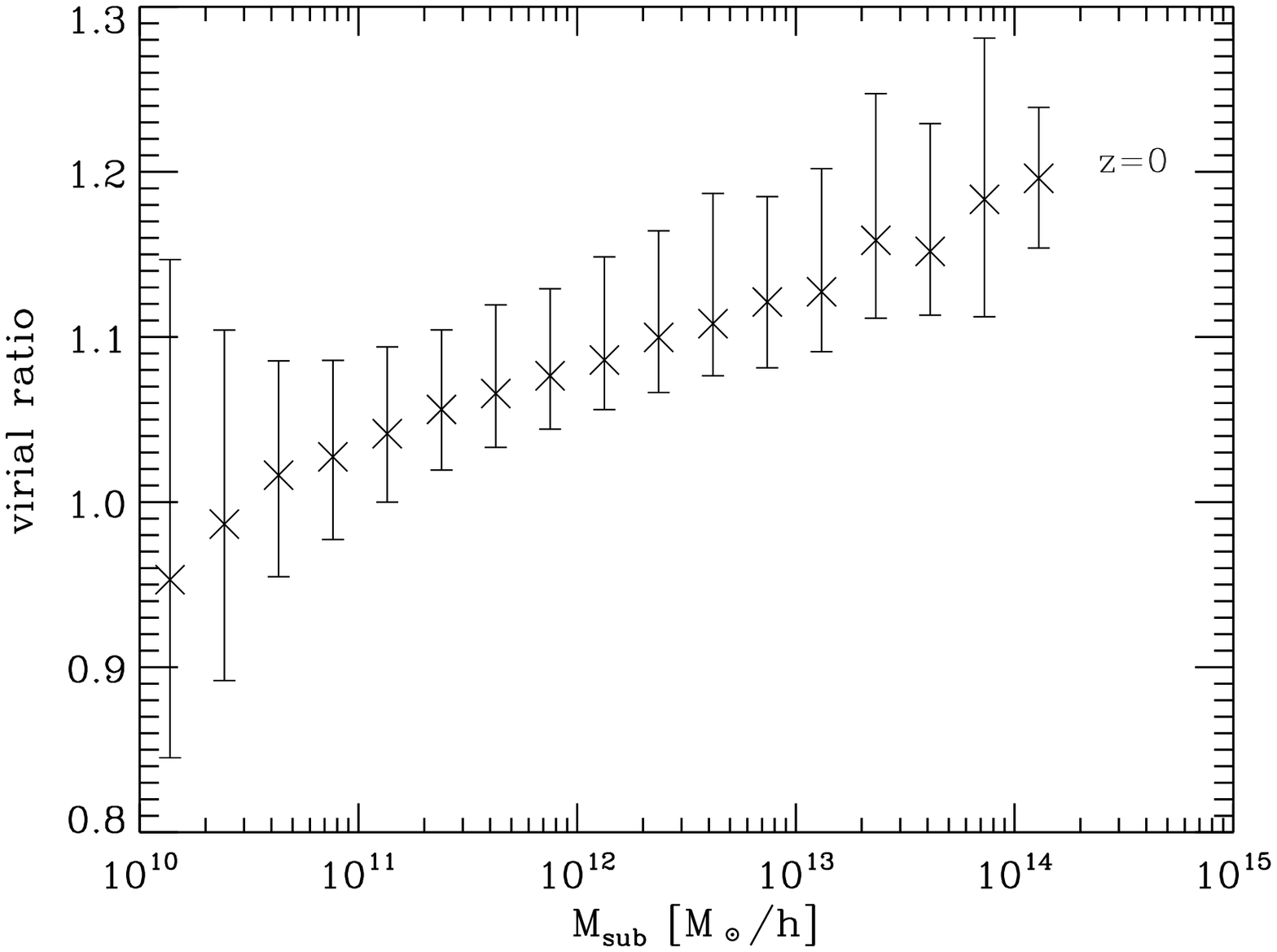}
  \caption{Distribution of the virial ratio in dark matter haloes of different mass at redshift $z = 0$. The virial ratio is defined such that it converges to unity at virial equilibrium. Crosses indicate the median value of the virial ratio, while the error bars indicate the 20th and 80th percentiles. The mass bins are spaced logarithmically, and symbols are only plotted for bins containing at least 10 dark matter haloes.}
\label{Mil-vr}
\end{figure}

In Fig.~\ref{Mil-tf}, we show the formation time of dark matter haloes found in the `Large' run, also identified by the method described in Section~\ref{mtree}. These results are consistent with the formation times as derived from the Millenium Simulation \citep{Har06}. Comparing Fig.~\ref{Mil-tf} with Fig.~\ref{tform}, a striking difference in the mass dependence of the formation time at the two different regimes is recognized: for dark matter structures present in the $z = 0$ simulations, the formation redshift decreases as a function of halo mass. In Fig.~\ref{Mil-vr}, the virial ratios of these haloes is presented (consistent with earlier findings of \citealt{Hetz06}). Comparing with Fig.~\ref{virial}, it is clear that dark matter structures identified at $z = 0$ are substantially closer to virial equilibrium compared to those at $z = 15$. In Fig.~\ref{Mil-med}, the median values of the sphericity parameter and triaxiality parameter are presented, which are consistent with results obtained from analyzing the Millenium Simulation \citep{Bet07}. Comparison with Fig.~\ref{stmed} shows that halos found in the larger simulation box at lower redshift are much more spherical, and less  prolate.

Overall, our results show good agreement with results found in the literature on large-scale structure formation at $z = 0$. This confirms the validity of the analysis methods we adopted in this work.

\section{Influence of particle numbers on calculating sphericity parameter.}

\begin{table*}
  \caption{Convergence study of our method to calculate halo shapes described in Section~\ref{sec_shape} using several different types of spheroids. Average and standard deviation is calculated from 100 realizations in each case. }
  \label{tab:abc}
  \begin{tabular}{lccc}\hline\hline

    real value  &  No. of particles   &     c/a   &  b/a  \\ \hline
    & 100  &   0.8135  (0.0575)  &   0.9083 (0.0484)  \\
    b/a=1,& 200  &   0.8649  (0.0467)  &   0.9345 (0.0327)  \\
    c/a=1 & 500                &    0.9109 (0.0296)   &   0.9558 (0.0228)  \\ 
    &   1000                   &    0.9402 (0.0217)   &   0.9708 (0.0184)  \\
    &   2000                   &    0.9562 (0.0147)   &   0.9782 (0.0131)  \\ \hline
    
    & 100                        &  0.6209 (0.0557)    &    0.8917 (0.0528) \\
    b/a=1,& 200                  &   0.6354 (0.0392)    &  0.9222 (0.0415)  \\
    c/a=2/3 &   500              &    0.6468 (0.0278)    &  0.9476 (0.0294) \\
    & 1000                       &    0.6532 (0.0196)    &  0.9655 (0.0206) \\
    &  2000                      &    0.6571 (0.0129)    &  0.9758 (0.0144) \\ \hline
    
    & 100                         &    0.3099  (0.0293)  &   0.8822 (0.0600)  \\
    b/a=1, & 200                  &    0.3175  (0.0200)  &   0.9152 (0.0472)  \\
    c/a=1/3 &   500               &    0.3234 (0.0125)   &   0.9423 (0.0286)  \\ 
    &   1000               &    0.3276 (0.0083)   &   0.9618 (0.0198)  \\
    &   2000               &    0.3292 (0.0063)   &   0.9734 (0.0144)  \\  \hline
    
    & 100                          &   0.3286  (0.0325)   &    0.6647 (0.0635) \\   
    b/a=2/3,  & 200                &   0.3313  (0.0227)   &    0.6658 (0.0476) \\
    c/a=1/3  &   500               &    0.3328 (0.0141)   &   0.6661 (0.0316)  \\ 
    &   1000               &    0.3339 (0.0092)   &   0.6662 (0.0214)  \\
    &   2000               &    0.3334 (0.0064)   &   0.6658 (0.0138)  \\ \hline
    
  \end{tabular}
\end{table*}

In order to test our algorithm to calculate the shape of dark matter haloes, we performed a series of controlled experiments; namely, we generated spherical distributions of particles with $\rho \propto r^{-2}$ from 100 different seeds for a range of particle numbers. Then, we deformed the sphere in $y$/$z$ direction with a ratio of $b/a = 1$, $c/a=2/3$, or $b/a = 1$, $c/a=1/3$, or $b/a = 2/3, c/a = 1/3$ and finally calculated halo shapes using the prescription described in Section~\ref{sec_shape}. We also measured the axis ratios of ideal spheres before applying any deformation. In this section, we adopt $b/a$ and $s$ (=$c/a$) to describe the shape of the spheroid, rather than $s$ and $T$ (=$(a^2 - b^2)/(a^2 - c^2)$), as in other parts of this paper. This is because triaxiality cannot be defined properly for an ideal sphere.

In Table~\ref{tab:abc}, the average over 100 realizations is shown with an error bar that indicates the standard deviation obtained from 100 realizations for different types of spheroids investigated here. There is a general trend that with increasing number of particles the measured axial ratios become ever closer to the real underlying value, and the standard deviation decreases.
 
It is shown here that in cases where the underlying values of $c/a$ or $b/a$ are unity, the real value and the estimated value do not match within standard deviations even with 2000 particles. Thus, our method tends to underestimate the axial ratios when they are close to unity. This tendency was already recognized earlier by \citet{Dub91}, who performed similar numerical experiments for their method to estimate halo shapes (their method is not exactly identical to the method we adopted in this work). 

\section{Resolution study of the non-dimensional spin parameter}
\label{discuss-l}

\begin{figure}
  \includegraphics[width=8cm]{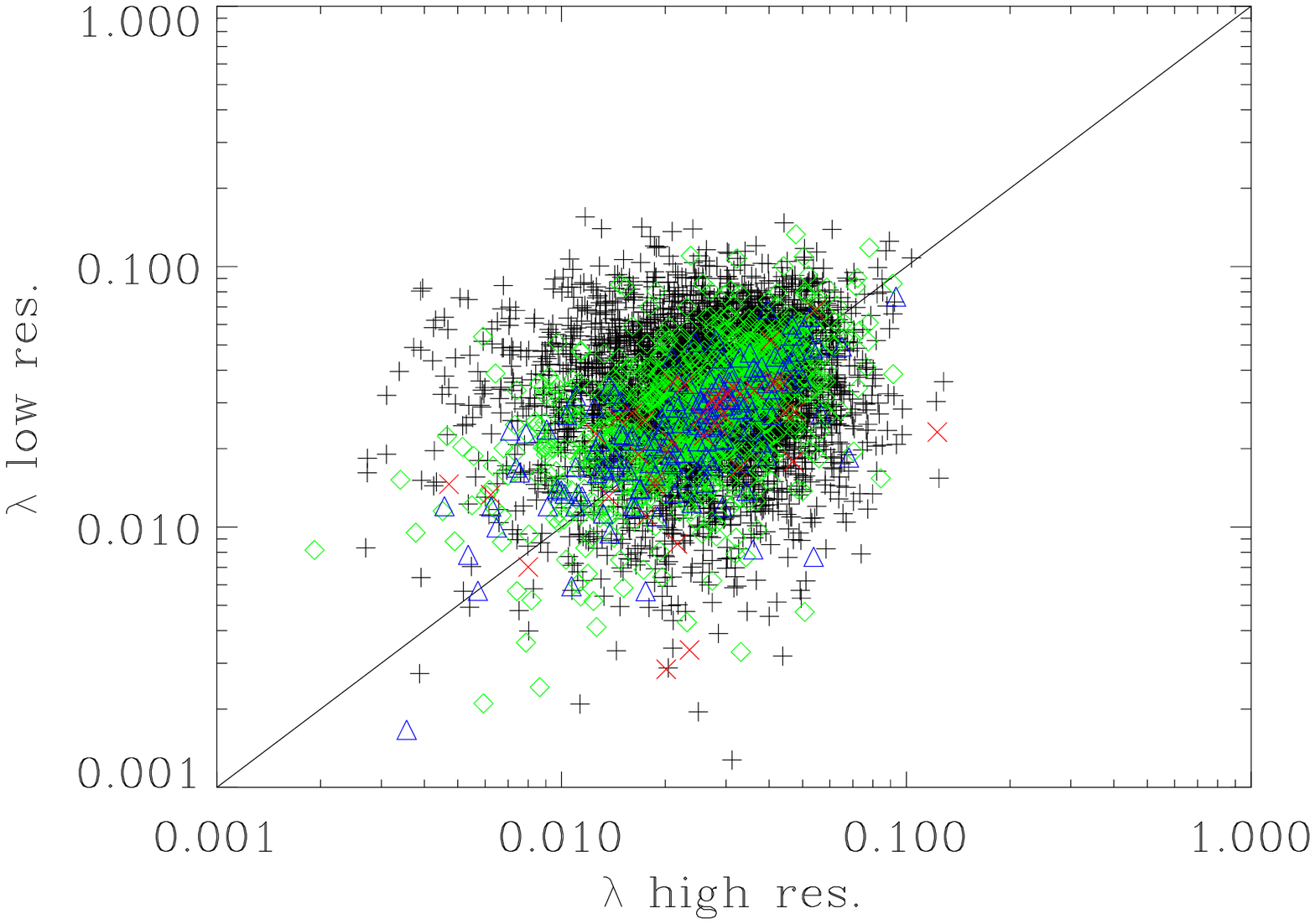}
  \caption{\MScolor{Non-dimensional spin parameter calculated from high-resolution ($2048^3$) simulation and lower-resolution ($512^3$) simulation. Black plus signs denote `main' subhalo with $<= $ 300 particles, green diamonds $>$ 300, $<=$ 1000 particles, blue triangles $>$ 1000, $<=$ 3000 particles, red crosses $>$ 3000 particles in the lower resolution simulation.}}
  \label{fig:lres}
\end{figure}

\MScolor{Since the non-dimensional spin parameter $\lambda$ is known to depend severely on how well the dark matter structure is resolved, we performed a resolution study by running a lower-resolution version of our `Small' run with $512^3$ particles adopting the same cosmological parameters and the same random seed for initial conditions. By matching dark matter haloes at $z=15$ in two different runs according to their positions, we compared the non-dimensional spin parameter. According to Fig.~\ref{fig:lres}, we find that haloes with $>$ 1000 particles in the lower resolution run, have relatively good convergence. Therefore, for the high-resolution production run, we show the spin parameter of only haloes with $>=$ 1000 particles, assuming that they have enough particles to capture the intrinsic angular momentum.}

\section{Non-dimensional spin parameter calculated for FOF haloes}
\label{discuss-fof}

\begin{figure*}
  \includegraphics[width=8cm]{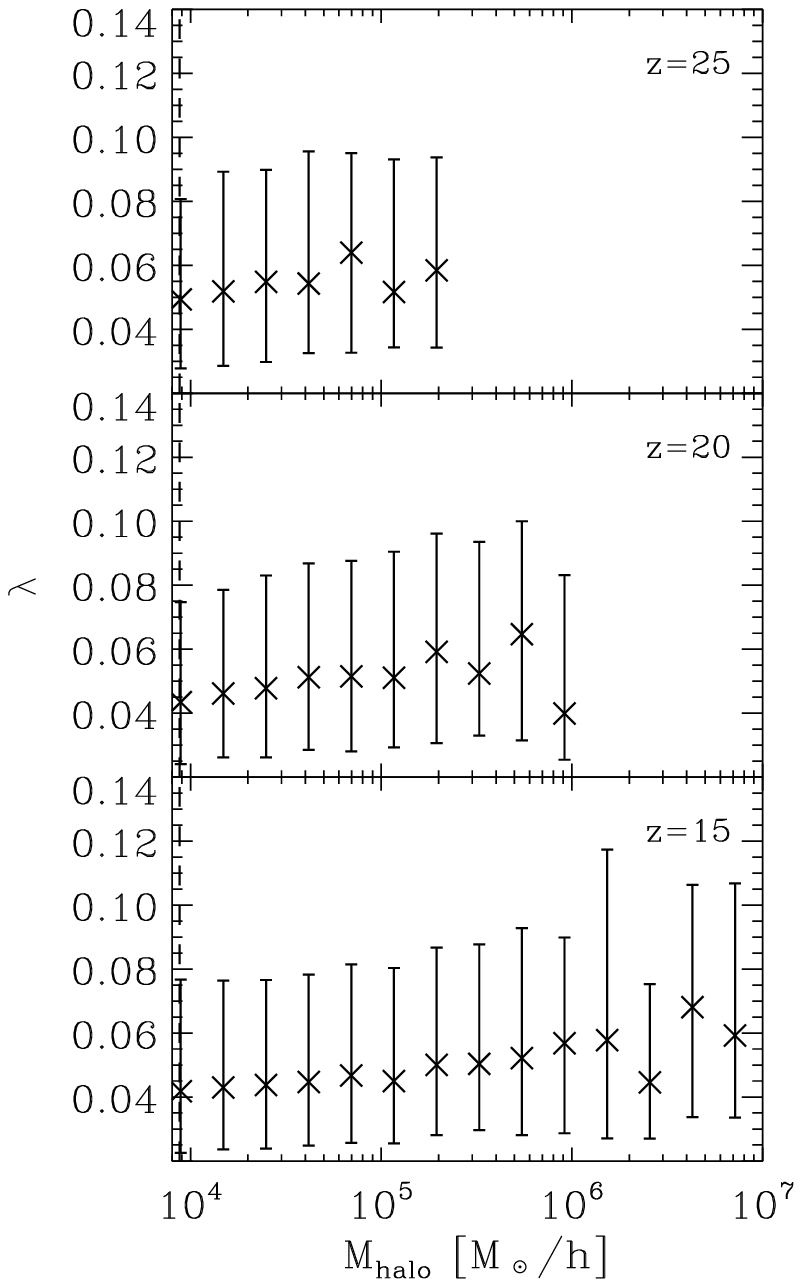}
  \caption{\MScolor{Distribution of the dimensionless spin parameter $\lambda$ in dark matter haloes of different mass at redshift $z \sim 25$ (top panel), $z \sim 20$ (middle panel), and $z \sim 15$ (bottom panel) calculated for {\bf FOF haloes}. Crosses indicate the median value of $\lambda$, while the error bars indicate the 20th and 80th percentiles. The dashed vertical lines represent lower mass limit of 1000 particles.}}
  \label{fig:FOF}
\end{figure*}

\MScolor{In Fig.~\ref{fig:FOF}, we present the non-dimensional spin parameter calculated for each {\bf FOF haloes} as a function of mass. Comparison with Fig.~\ref{spin15} shows that FOF gives a positive correlation between spin and halo mass, which is not present for substructures identified by {\small SUBFIND}.}

\section{Density slice through the whole box.}

\begin{figure*}
  \begin{center}
    \begin{tabular}{ll}
      \includegraphics[width=8cm]{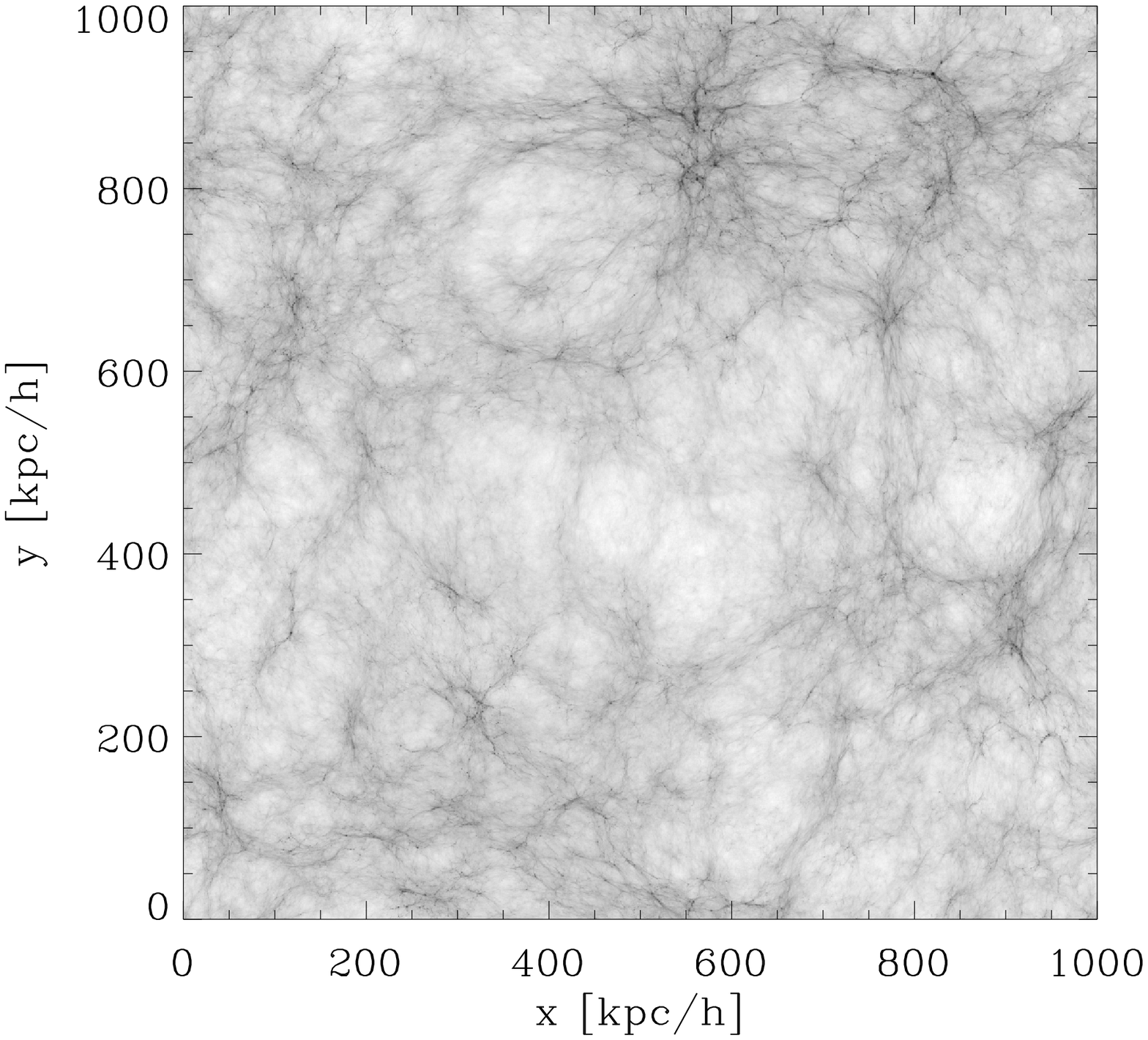} &
      \includegraphics[width=8cm]{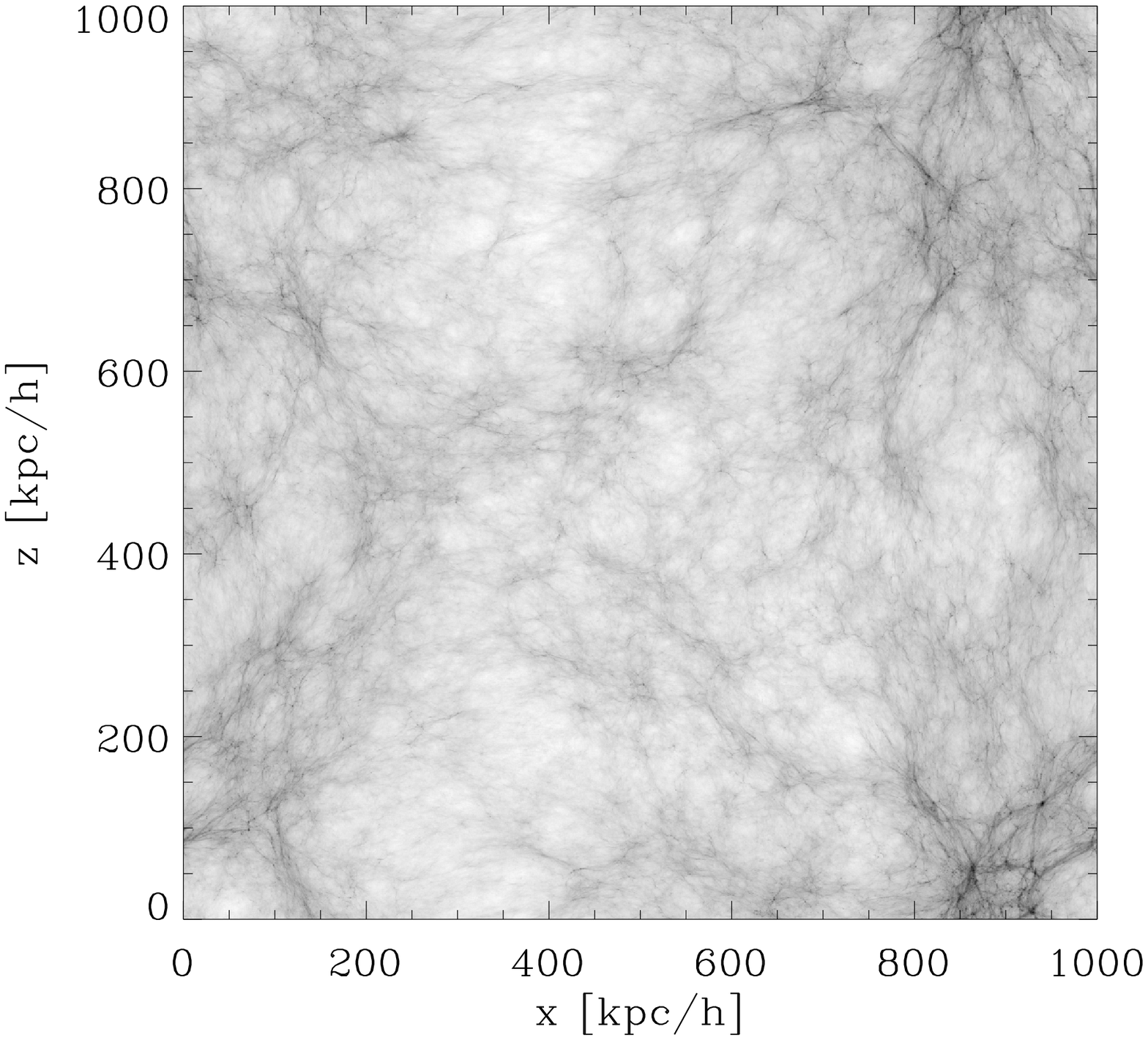} \\
    \end{tabular}
  \end{center}
  \caption{A density projection onto the $x$-$y$ ($x$-$z$) plane through the
 center of the box with thickness one-fifth of the simulation box size
 for our `Small' run at $z = 15$.}
  \label{x-y}
\end{figure*}

\begin{figure*}
  \begin{center}
    \begin{tabular}{cc}
      \includegraphics[width=8cm]{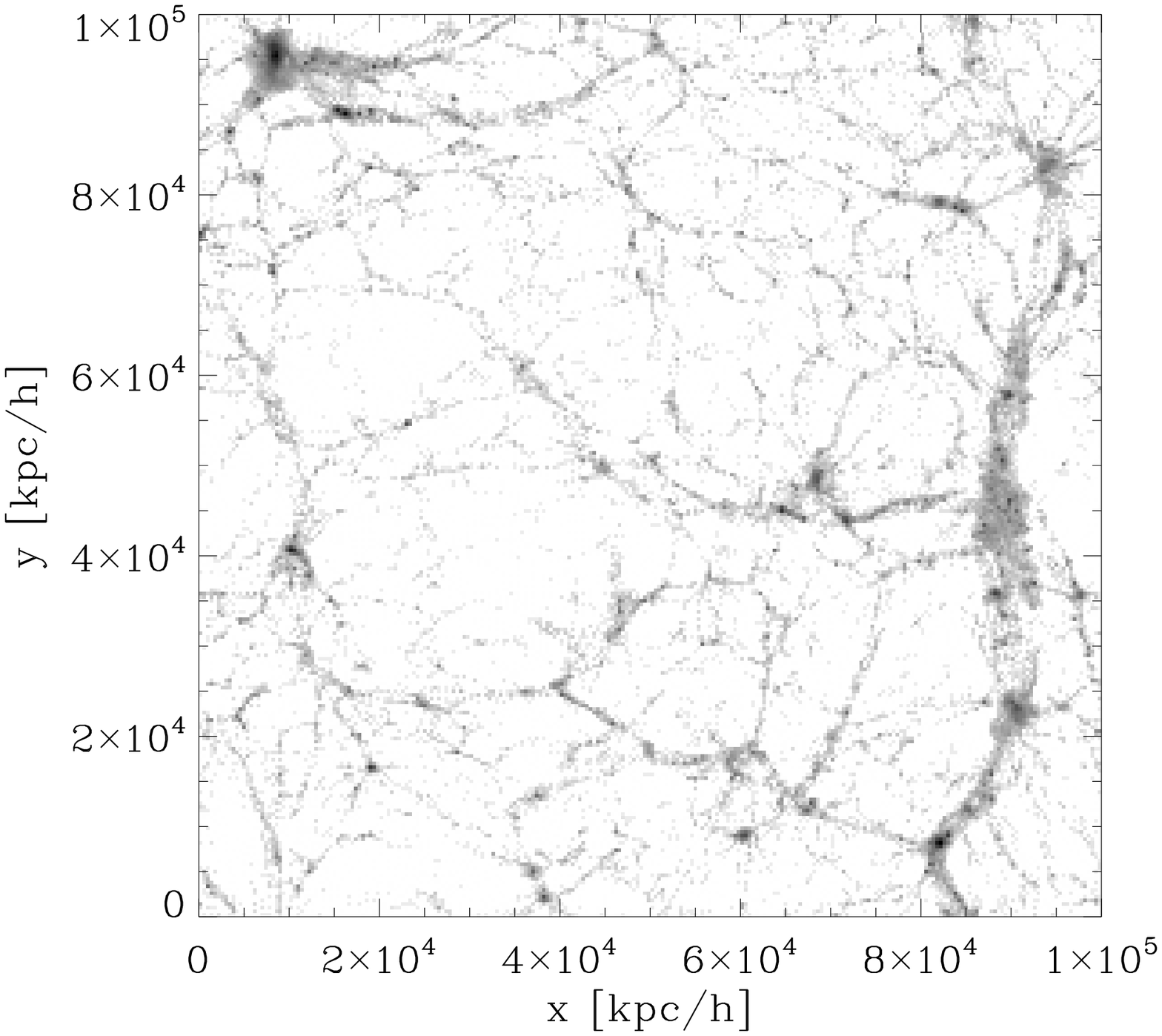} &
      \includegraphics[width=8cm]{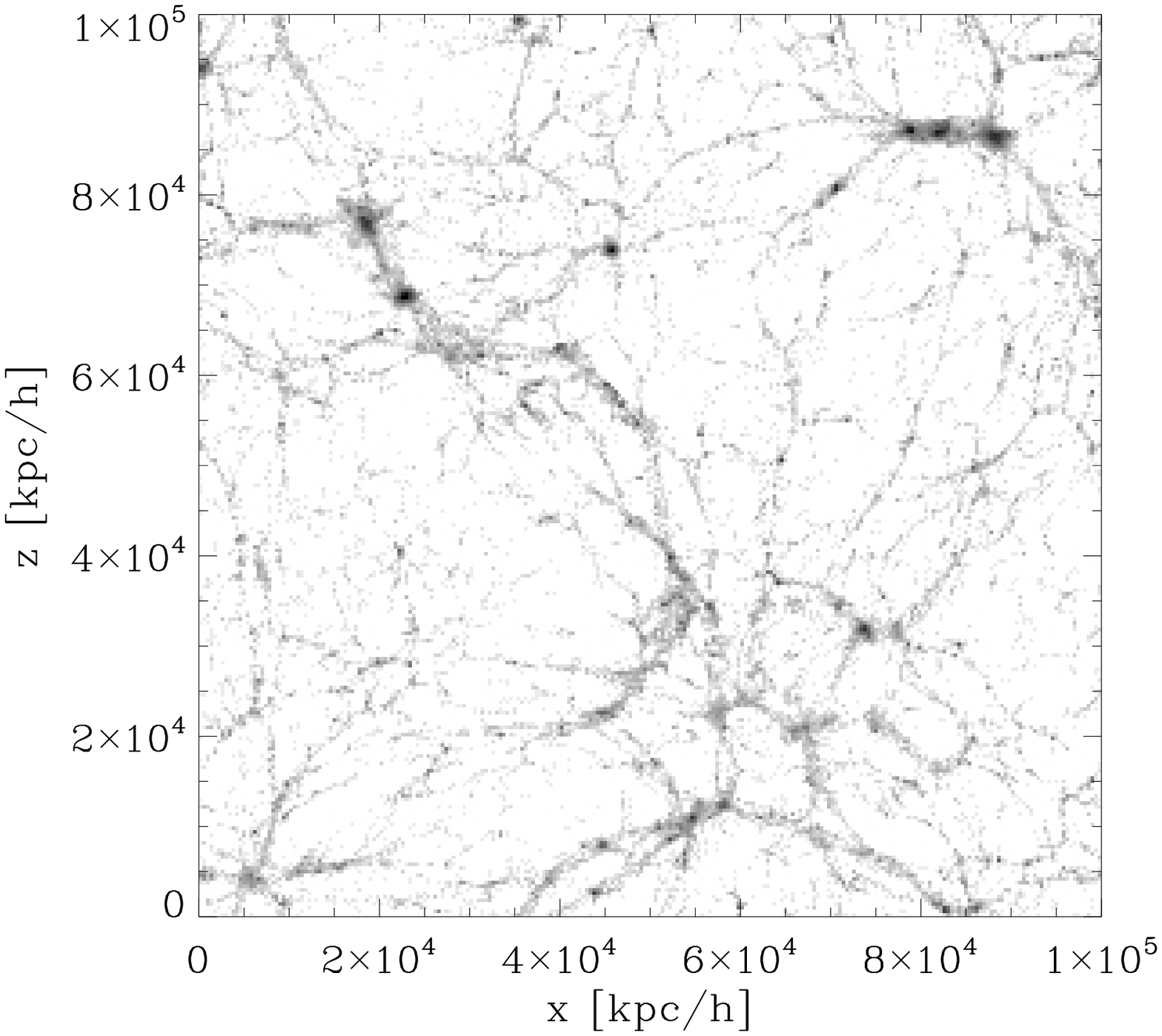} \\
    \end{tabular}
  \end{center}
  \caption{A density projection onto the $x$-$y$ ($x$-$z$) plane through the
 center of box with thickness one-fifth of the simulation box size for
our `Large' run at $z = 0$.}
  \label{x-y_0}
\end{figure*}

In Fig.~\ref{x-y} and \ref{x-y_0}, we present the density projection
onto the $x$-$y$ and $x$-$z$ planes from our `Small' and `Large' runs,
respectively, in order to illustrate the differences in structure
formation at different scales.

\label{lastpage}

\end{document}